\documentclass[acmsmall,screen]{acmart}

\AtBeginDocument{%
  }

\setcopyright{acmlicensed}
\copyrightyear{2025}
\acmYear{2025}
\acmDOI{XXXXXXX.XXXXXXX}
\acmJournal{TOSEM}
\acmVolume{1}
\acmNumber{1}
\acmArticle{111}
\acmMonth{12}
\acmISBN{978-1-4503-XXXX-X/2018/06}

\usepackage{amsmath}
\usepackage[ruled,vlined,linesnumbered]{algorithm2e}
\usepackage{multirow}
\usepackage{bigstrut}

\usepackage{amssymb}
\usepackage{newtxmath}
\usepackage{amsthm}
\usepackage{booktabs}
\usepackage{graphicx}
\usepackage{pdfpages}
\usepackage{mdframed}
\usepackage{adjustbox}

\usepackage{float}

\usepackage[justification=centering]{caption}

\usepackage{colortbl}

\usepackage{tikz}

\newcommand{\ourSol}{\textsc{Vital}\xspace}
\newcommand{\chopper}{\texttt{Chopper}\xspace}
\newcommand{\ccured}{\texttt{CCured}\xspace}
\newcommand{\ourSolNoSim}{\textsc{Vital($\lnot$Sim)}\xspace}

\newcommand{\ourSolNoExp}{\textsc{Vital($\lnot$Exp)}\xspace}

\newcommand{\ourSolNoSimOpt}{\textsc{Vital($\lnot$Sopt)}\xspace}

\newcommand{\ourSolOptTwoF}{\textsc{Vital(OPT2-500)}\xspace}
\newcommand{\ourSolOptTwoO}{\textsc{Vital(OPT2-1000)}\xspace}
\newcommand{\ourSolOptTwoT}{\textsc{Vital(OPT2-2000)}\xspace}

\newcommand{\cgs}{\textsc{Cgs}\xspace}
\newcommand{\empc}{\textsc{Empc}\xspace}
\newcommand{\featmaker}{\textsc{FeatMaker}\xspace}

\usepackage{array} 
\usepackage{framed}
\usepackage{booktabs}
\usepackage{subfigure}
\usepackage[justification=centering]{caption}
\usepackage[figuresright]{rotating}
\usepackage{listings}
\usepackage{xcolor}
\usepackage[ruled,vlined,linesnumbered]{algorithm2e}

\SetCommentSty{mycommfont}

\usepackage{comment}
\usepackage[T1]{fontenc} 
\usepackage{array}
\usepackage{url}
\usepackage[shortlabels]{enumitem}
\usepackage{pifont}

\usepackage{flushend}
\usepackage{ragged2e}
\usepackage{lipsum}

\definecolor{mygreen}{rgb}{0,0.5,0}
\definecolor{myblue}{rgb}{0,0,1}
\definecolor{mymauve}{rgb}{0.58,0,0.82}
\definecolor{myblack}{rgb}{0.24,0.17,0.12}
\definecolor{awesome}{rgb}{1.0, 0.13, 0.32}
\usepackage{xcolor}
\usepackage{tikz}

\usepackage[normalem]{ulem} 
\newcommand\hl{\bgroup\markoverwith
	{\textcolor{lightgray}{\rule[-.5ex]{2pt}{2.5ex}}}\ULon}

\usepackage{hyperref}
\definecolor{light-gray}{gray}{0.80}
\usepackage{varwidth}
\usepackage{threeparttable}
\usepackage{xcolor}

\newcommand{\red}[1]{\textcolor{red}{#1}} %

\newcommand{\rev}[1]{\textcolor{black}{#1}} %

\usepackage[most]{tcolorbox}
\newtcolorbox{resultbox}{
  colback=gray!10,     
  colframe=gray!70,    
  boxrule=0.8pt,       
  arc=4pt,             
  left=6pt, right=6pt, top=6pt, bottom=6pt, 
  enhanced
}

\newcommand{\cc}[1]{{#1}} 

\begin{document}
\title{\ourSol: Vulnerability-Oriented Symbolic Execution via Type-Unsafe Pointer-Guided Monte Carlo Tree Search}

        \author{Haoxin Tu}
        \authornote{Partial work was done when Haoxin was visiting the MPI Software Security Group led by Prof. Marcel Böhme.}
	\affiliation{%
	\institution{Singapore Management University}
	\country{Singapore}
	}
	\email{haoxintu@gmail.com}
	
	\author{Lingxiao Jiang}
	\affiliation{%
	\institution{Singapore Management University}
	\country{Singapore}
	}
	\email{lxjiang@smu.edu.sg}
        
        \author{Marcel Böhme}
	\affiliation{%
	\institution{Max Planck Institute for Security and Privacy}
	\country{Germany}
	}
	\email{marcel.boehme@acm.org}

\renewcommand{\shortauthors}{Haoxin Tu, Lingxiao Jiang, and Marcel Böhme}

\begin{abstract}
  How do we find new memory safety bugs effectively when navigating a symbolic execution tree that suffers from the well-known path explosion challenge?
Existing solutions either adopt path search heuristics to maximize coverage rate or chopped symbolic execution to skip uninteresting code (i.e., manually labeled as vulnerability-unrelated) during path exploration.  
However, most existing search heuristics are not vulnerability-oriented, and manual labeling of irrelevant code-to-be-skipped relies heavily on prior expert knowledge, making it hard to detect vulnerabilities effectively in practice.

This paper proposes \ourSol, a new vulnerability-oriented path exploration for symbolic execution with two innovations.
First, a new indicator (i.e., {\it type-unsafe} pointers) is suggested to approximate vulnerable paths.
A pointer that is \emph{type-unsafe} cannot be statically proven to be safely dereferenced without memory corruption.
Our key hypothesis is that a path with more {\it type-unsafe} pointers is more likely to be vulnerable.
Second, a new {\it type-unsafe} pointer-guided Monte Carlo Tree Search algorithm is implemented to guide the path exploration towards the areas that contain more {\it unsafe} pointers, aiming to increase the likelihood of detecting vulnerabilities.
We built \ourSol on top of KLEE and compared it with existing path searching strategies and chopped symbolic execution.
In the former, the results demonstrate that \ourSol could cover up to 90.03\% more unsafe pointers and detect up to 57.14\% more unique memory errors. 
In the latter, the results show that \ourSol could achieve a speedup of up to 30x execution time and a reduction of up to 20x memory consumption to detect known vulnerabilities without prior expert knowledge automatically. 
In practice, \ourSol also detected one previously unknown vulnerability  (a new CVE ID is assigned), which has been fixed by developers.

\end{abstract}

\ccsdesc[500]{Software and its engineering~Software testing and debugging}

\keywords{Reliability, Security, Program analysis, Symbolic execution.}

\maketitle

\section{Introduction}  \label{mcts::sec:introduction}

Memory unsafety is the leading cause of vulnerabilities in complex software systems \cite{0day,memory-safety-chromum,memory-safety-google,blog-microsoft}. 
For example, based on a report from Microsoft, more than 70\% of security bugs have been memory safety problems in the last 12 years \cite{blog-microsoft}. 
To prevent memory errors, many advanced static/dynamic/symbolic analysis-based approaches are devoted to detecting such errors automatically. Among them, \emph{symbolic execution} is considered one of the promising program analysis techniques \cite{angr,symloc,klee}. 
The key idea of symbolic execution is to simulate program executions with symbolic inputs and generate test cases by solving path constraints collected during execution.
Taking advantage of the soundness of test case generation, symbolic execution has also been applied in many other areas, such as programming language \cite{symcc, computing-summary, state-merging} and security \cite{qsym,aeg,mayhem,cottontail,coclimic}. 

One of the key challenges in symbolic execution is path explosion, where even a small program can produce a vast execution tree \cite{survey-se2018}.
Two main alternatives have been proposed to alleviate this challenge.
The first alternative is to steer the exploration of the path to maximize the rate at which the new code is covered using various search heuristics. Notably, the symbolic execution KLEE \cite{klee} supports random, breadth-first ({\tt bfs}), depth-first search ({\tt dfs}), and other coverage-guided heuristics.
A recent work, CBC \cite{yi2024compatible}, proposes to use compatible branch coverage-driven path exploration for symbolic execution.
\rev{\cgs \cite{cgs-icse24}, \featmaker \cite{featMaker-fse24}, and \empc \cite{empc-sp25} are the most recent works that propose new path search heuristics to improve code coverage.}
However, existing studies show that achieving the best coverage does not necessarily mean that the largest number of vulnerabilities can be detected \cite{bohme2022reliability,chen2020savior}.

Another solution is to skip the symbolic execution of certain code that is manually labeled as not related to the vulnerability using \emph{chopped symbolic execution} \cite{trabish2018chopped}. 
However, setting up chopped execution requires prior expert knowledge of the program under test and intensive manual effort to decide which functions to skip.
For example, to successfully detect a vulnerability (CVE-2015-2806) in the {\tt libtasn1} library, users need to locate four specific functions and two lines to skip\footnote{Manually defined option used in \chopper: ``{\it --skip=\_\_bb0,\_\_bb1,\_asn1\_str\_cat:403/404,asn1\_delete\_structure}''.}. Since the library includes more than 20{,}000 lines of code, locating these functions/lines may require expert knowledge and involve intensive human efforts.
Also, it is extremely difficult to use it to find new vulnerabilities, as we do not know where the vulnerabilities are until they are caught.
Another issue with chopped executions is their performance. It consumes significant memory when switching between skipped and normal execution (see more details in Section~\ref{mcts::sec:evaluation:rq2:memory-usage}).
Motivated by addressing the above limitations, we aim to investigate the following research question:
{\it How can we automatically conduct vulnerability-oriented path exploration to detect new vulnerabilities without relying on prior expert knowledge?}

Conducting vulnerability-oriented path exploration is non-trivial and presents challenges for at least the following two reasons. 
First, we should decide which indicators can effectively approximate the vulnerability of a path. To our knowledge, no existing unified metrics could be used to indicate a path containing memory errors (e.g., {\it out-of-bounds}).
Therefore, it is challenging to find a suitable indicator to represent vulnerable paths.
Second, it is also difficult to effectively leverage indicators to guide path exploration. 
To make path exploration more effective, we argue that a promising path search strategy should maintain a good trade-off between exploiting the paths that have already been executed in the past and exploring the paths that will be executed in the future.
However, existing search heuristics do not fully take advantage of past execution, which degrades the possibility of exploring likely vulnerable paths.

\begin{figure}[t]
	\centering
	\includegraphics[width=0.5\linewidth]{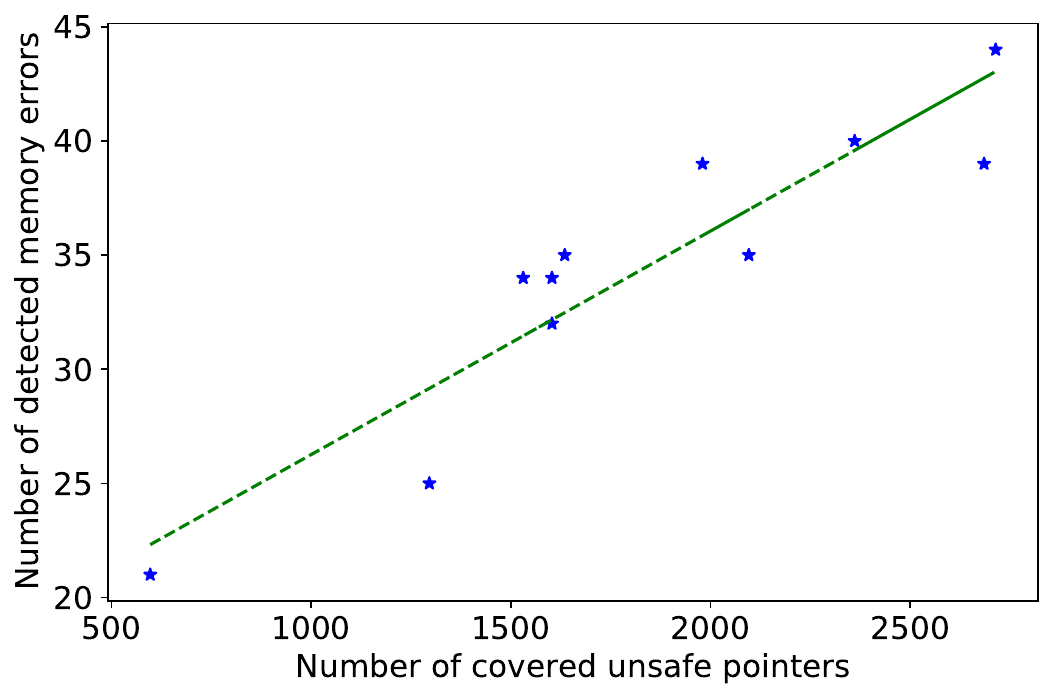}
	\vspace{-1em}
	\caption{Correlation between unsafe pointers and memory errors (Pearson's coefficient \cite{cohen2009pearson}: 0.924)}
	\label{mcts::fig:correlation}
\end{figure}

In this paper, we propose \ourSol\footnote{The name also reflects our aim for exploring {\it vital} program paths due to path explosion.}, a new {\bf V}ulnerability-or{\bf I}en{\bf T}ed p{\bf A}th exp{\bf L}oration strategy for symbolic execution via type-unsafe pointer-guided Monte Carlo Tree Search (MCTS).
Our core insight is that spatial memory safety errors (e.g., {\it out-of-bounds} memory accessing errors) can {\it only} happen when dereferencing {\it type-unsafe} pointers, i.e., pointers that cannot be statically proven to be memory safe \cite{ccured-popl2002,dataGuard,szekeres2013sok}. As shown in Figure \ref{mcts::fig:correlation}, where the data is from in our experiments (in Section \ref{mcts::sec:evaluation:rq1}), we found that an increase in the number of unsafe pointers exercised is directly related to an increasing number of memory errors, with a strong positive Pearson coefficient 0.93 \cite{cohen2009pearson}.
Hence, we suggest maximizing the number of unsafe pointers during path exploration to address the first challenge of approximating vulnerability. 
To address the second challenge of effective search within the execution tree, we drive a new pointer-guided MCTS that effectively balances the path exploration and exploitation to maximize the number of covered unsafe pointers.

We implement \ourSol on top of KLEE \cite{klee} and conduct extensive experiments to compare \ourSol with existing search strategies and chopped symbolic execution.
We run \ourSol against six path exploration strategies in KLEE and a recently proposed strategy, CBC \cite{yi2024compatible} on the GNU {\tt Coreutils}, and the results demonstrate that \ourSol could cover up to 90.03\% more unsafe pointers and detect up to 57.14\% more unique memory errors. 
Further, we run \ourSol against KLEE \cite{klee}, \chopper \cite{trabish2018chopped}, and CBC \cite{yi2024compatible} over six known CVE vulnerabilities, and the results show that \ourSol could achieve a speedup of up to 30x execution time and a reduction of up to 20x memory consumption to detect all of them automatically without prior expert knowledge. 
\cc{\ourSol also detected one previously unknown vulnerability missed by 
other approaches. 
It has been confirmed and fixed by developers, with a new CVE ID (i.e., CVE-2025-3198) assigned.
}

\smallskip
In summary, this paper makes the following contributions.
\begin{itemize}[leftmargin=1em,itemsep=0.1cm]
    \item To our knowledge, \ourSol is the first work performing vulnerability-oriented path exploration for symbolic execution towards effective and efficient vulnerability detection.
    \item We suggest a new indicator (i.e., the number of type-unsafe pointers) to approximate the vulnerability (or vulnerability-proneness) of a program path and utilize a new type-unsafe pointer-guided Monte Carlo Tree Search algorithm to navigate smart vulnerability-oriented path searches.
    \item Extensive experiments demonstrate the superior performance of \ourSol in terms of unsafe pointer coverage and memory errors/vulnerability detection compared to state-of-the-art approaches. In practice, \ourSol also detected a previously unknown vulnerability, indicating its practical vulnerability detection capability.
    \item We publish the replication package online at {\it \textcolor{mymauve}{\url{https://github.com/haoxintu/Vital-SE}}}, to foster practical symbolic execution for vulnerability detection.
\end{itemize}

{\bf Organizations.} Section \ref{mcts::sec:motivation} gives the background and motivation. 
Section \ref{mcts::sec:approach} describes the design of \ourSol.
Section \ref{mcts::sec:evaluation} presents our evaluation results. Section \ref{mcts::sec:discussion} discusses the overhead of type inference, the impact of different configurations, and threats to validity.
Sections \ref{mcts::sec:related-work} and \ref{mcts::sec:conclusion} describe related works and conclude with future work. We also provide extra discussions and experimental results in the Appendix in the Supplementary Material. 

\section{Background and Motivation}  \label{mcts::sec:motivation}

\subsection{Background}  \label{mcts::sec:background}

\subsubsection{Symbolic Execution and Path Exploration}  \label{mcts::sec:back-path-exploration} 
Symbolic execution is a program analysis technique that analyzes a test program by feeding the program with symbolic inputs \cite{boosting-se,survey-se2018}.
During symbolic execution, path constraints are collected, and corresponding test cases will be generated by off-the-shelf constraint solvers (e.g., STP \cite{stp} or Z3 \cite{z3}).
The main activity in symbolic execution is to consistently select a path (or state) to explore and analyze an instruction one time until no state remains or a given timeout is reached.
Notably, the widely used KLEE symbolic execution engine, the representative search strategies include {\tt bfs}, {\tt dfs}, {\tt random}, code coverage-guided (i.e., {\tt nurs:covnew}), and instruction coverage-guided (i.e., {\tt nurs:md2u} and {\tt nurs:icnt}).
\rev{Recent work CBC \cite{yi2024compatible}, \cgs \cite{cgs-icse24}, \featmaker \cite{featMaker-fse24}, and \empc \cite{empc-sp25} also propose new search strategies to improve path exploration in symbolic execution.}

\subsubsection{Memory Safety and Type Inference}  \label{mcts::sec:back-memory-safety}
Existing memory safety vulnerabilities fall mainly into two main categories: {\it spatial} and {\it temporal} memory safety vulnerabilities \cite{dataGuard,symloc}. 
The first ones occur when pointers reference addresses outside the legitimate bounds (e.g., {\it buffer overflow}), and {\it temporal} memory safety issues arise from the use of pointers outside of their live period (e.g., {\it use-after-free}).
Previous studies \cite{nagarakatte2009softbound,bogo-asplos19} show that serious vulnerabilities caused by violating spatial safety are well known in the community.
For example, more than 50\% of recent CERT warnings are due to breaches of spatial safety \cite{wagner2000first}.
Furthermore, compared to temporal safety issues, spatial safety vulnerabilities can be exploited by many mature techniques, such as return-oriented programming \cite{prandini2012return} and code reuse attacks \cite{bletsch2011jump,checkoway2010return}.
The above fact indicates that designing advanced solutions to keep spatial memory safe is of critical importance.

To enforce complete spatial safety, the only way is to keep track of the pointer bounds (the lowest and highest valid address to which it can point) \cite{szekeres2013sok}.
Integrating a type inference system by static analysis into a type-unsafe programming language (e.g., C) is a well-established strategy for keeping track of pointer bounds \cite{elliott2018checked,dataGuard} to prove memory safety.
For example, \ccured \cite{ccured-popl2002} statically infers pointer types into three categories: {\it SAFE}, Sequence ({\it SEQ}), and Dynamic ({\it DYN}).
The {\it SAFE} pointers are the dominant portion of the program (e.g., 92.29\% of the pointers are safe reported by a recent work \cite{dataGuard}) that can be proved to be free of memory errors \cite{ccured-popl2002}.
The {\it SEQ} pointers require extra bound checking, and {\it DYN} pointers need to perform run-time checks to ensure memory safety. 
Since spatial memory error can only occur within {\it SEQ} or {\it DYN} pointer categories, we refer to {\it SEQ} or {\it DYN} pointers as {\it type-unsafe} or {\it unsafe} pointers in this paper. Taking the code from Figure \ref{mcts::fig:motivating-example}(a) for example, the pointers {\tt p.y} (Line 9), {\tt p.z} (Line 16), and {\tt p.y} (Line 20) are used in pointer arithmetic, which are classified as {\it unsafe} ({\it SEQ}) pointers. Such classification results can be obtained with mature type inference tools such as Ccured \cite{dataGuard}.

\subsubsection{Monte Carlo Tree Search (MCTS)} \label{mcts::sec:back-mcts}
MCTS is a heuristic search algorithm to solve decision-making problems, especially those in games (e.g., AlphaGo) \cite{mcts2012survey,luckow2018monte}. The fundamental idea behind MCTS is to use \cc{reward-weighted} randomness to simulate decision sequences and then to use the results of these simulations to decide the most promising move. 
It excels at balancing between exploring new moves in the future and exploiting known good moves in the past.

Technically, the MCTS consists of the following four steps (see Section 3.1 in \cite{mcts2012survey}):
\begin{itemize}[leftmargin=1em,itemsep=0.1cm]
    \item {\it Selection}: starting from the root node, select successive child nodes that are already in the search tree according to the {\it tree policy} down to an expandable node that has unvisited children.
    \item {\it Expansion}: expand the unvisited node into the search tree.
    \item {\it Simulation}: perform a playout from the expanded node governed by a {\it simulation policy}. The termination of the simulation yields results (i.e., rewards) based on a reward function.
    \item {\it Backpropagation}: update the information stored in the nodes on the path from the expanded node to the root node with the calculated reward. This also includes updating the visit counts.
\end{itemize}

Note that the {\it tree policy} (including node selection and expansion) and {\it simulation policy} are two key factors in conducting effective MCTS applications. Since then, a large number of enhancements have been proposed to facilitate the capabilities of MCTS \cite{mcts2012survey,swiechowski2023monte}.

\subsection{Motivating Example}

\begin{figure}[t]
\vspace{-1em}
\centering
\subfigure[Code Example]{
\begin{minipage}[t]{0.47\linewidth}
\centering
\includegraphics[width=0.6\linewidth]{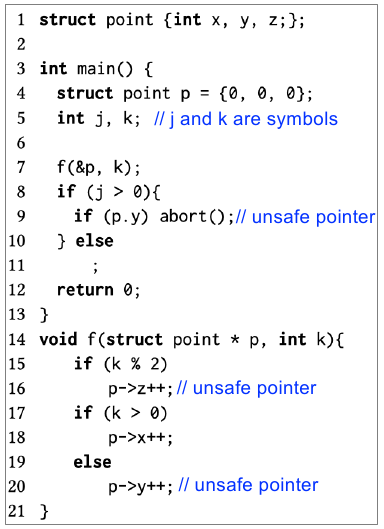}
\vspace{-1em}
\end{minipage}
}
\subfigure[Execution tree of code example]{	
\begin{minipage}[t]{0.47\linewidth}
\centering
\includegraphics[width=0.99\linewidth]{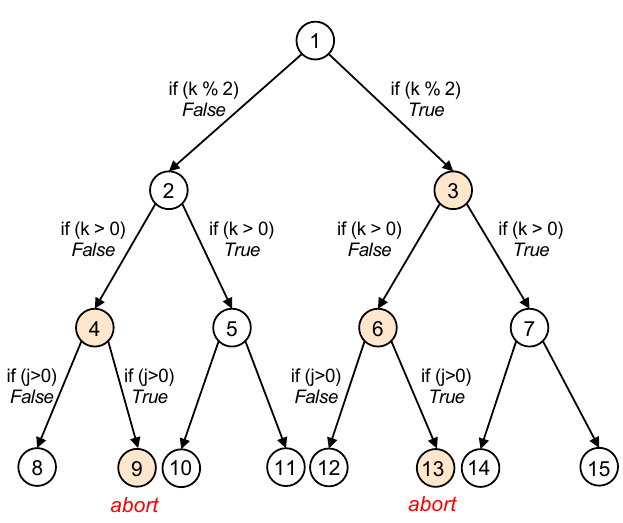}
\vspace{-1em}

\end{minipage}	
}
\vspace{-1em}
\caption{Motivating example (the code example is adapted from \chopper \cite{trabish2018chopped})}

\label{mcts::fig:motivating-example}
\end{figure}

To illustrate the motivation behind \ourSol, we will use a simple example shown in Figure \ref{mcts::fig:motivating-example}(a) (adapted from \chopper \cite{trabish2018chopped}) to show the limitations of the existing solution and the advantage of \ourSol. 
Figure \ref{mcts::fig:motivating-example}(b) represents the corresponding process tree, where the arrow denotes the execution flow with branch conditions, and each circle represents an execution state. 
The colored nodes mean they include unsafe pointers, where the pointers are code blocks that are unique to two forked states. 
For example, when the executor forks at Line 17, the code in Lines 18 and 19 is two unique code blocks for the two forked states.
The eight leaf nodes are terminal states, while the others are internal states.  
Assuming that we are trying to answer the following question:
{\it How can we efficiently trigger the problematic {\tt abort} failure at Line 9?}

\smallskip
{\bf Direction 1: Path Search Heuristics.}
Two commonly used search strategies are {\tt bfs} and {\tt dfs}, both of which apply a fixed order to explore all program paths in the following ranked order, where the leaf nodes represent the termination states (the ``->'' represents the order of terminated states):
\begin{itemize}[leftmargin=1em,nosep]
    \item {\tt bfs}:
    \tikz[baseline=(char.base)] \node[shape=circle,draw,fill=white!20,inner sep=1pt, minimum size=4mm, font= \footnotesize] (char) {15}; ->
    \tikz[baseline=(char.base)] \node[shape=circle,draw,fill=white!20,inner sep=1pt, minimum size=4mm, font= \footnotesize] (char) {14}; ->
    \tikz[baseline=(char.base)] \node[shape=circle,draw,fill=orange!20,inner sep=1pt, minimum size=4mm, font= \footnotesize] (char) {13}; ->
    \tikz[baseline=(char.base)] \node[shape=circle,draw,fill=white!20,inner sep=1pt, minimum size=4mm, font=\footnotesize] (char) {12}; ->
    \tikz[baseline=(char.base)] \node[shape=circle,draw,fill=white!20,inner sep=1pt, minimum size=4mm, font=\footnotesize] (char) {11}; ->
    \tikz[baseline=(char.base)] \node[shape=circle,draw,fill=white!20,inner sep=1pt, minimum size=4mm, font=\footnotesize] (char) {10}; ->
    \tikz[baseline=(char.base)] \node[shape=circle,draw,fill=orange!20,inner sep=1pt, minimum size=4mm, font=\footnotesize] (char) {9}; ->
    \tikz[baseline=(char.base)] \node[shape=circle,draw,fill=white!20,inner sep=1pt, minimum size=4mm, font=\footnotesize] (char) {8};
    \item {\tt dfs}:
    \tikz[baseline=(char.base)] \node[shape=circle,draw,fill=white!20,inner sep=1pt, minimum size=4mm, font=\footnotesize] (char) {8}; ->
    \tikz[baseline=(char.base)] \node[shape=circle,draw,fill=orange!20,inner sep=1pt, minimum size=4mm, font=\footnotesize] (char) {9}; ->
    \tikz[baseline=(char.base)] \node[shape=circle,draw,fill=white!20,inner sep=1pt, minimum size=4mm, font= \footnotesize] (char) {10}; ->
    \tikz[baseline=(char.base)] \node[shape=circle,draw,fill=white!20,inner sep=1pt, minimum size=4mm, font= \footnotesize] (char) {11}; ->
    \tikz[baseline=(char.base)] \node[shape=circle,draw,fill=white!20,inner sep=1pt, minimum size=4mm, font= \footnotesize] (char) {12}; ->
    \tikz[baseline=(char.base)] \node[shape=circle,draw,fill=orange!20,inner sep=1pt, minimum size=4mm, font= \footnotesize] (char) {13}; ->
    \tikz[baseline=(char.base)] \node[shape=circle,draw,fill=white!20,inner sep=1pt, minimum size=4mm, font= \footnotesize] (char) {14}; ->
    \tikz[baseline=(char.base)] \node[shape=circle,draw,fill=white!20,inner sep=1pt, minimum size=4mm, font= \footnotesize] (char) {15};
\end{itemize}

We can observe that neither of them can explore (or rank) the path containing the {\tt abort} failure at the top-1 position, so they might not be efficient.
A random search might first find the {\tt abort} path, but the result is unreliable.
Other search heuristics, such as coverage-guided (e.g., CBC \cite{yi2024compatible}), are good at covering new code but are still likely to miss certain errors because the best coverage has no statistical relation with the largest number of bugs \cite{bohme2022reliability}.

\smallskip
{\bf Direction 2: Chopped Symbolic Execution.} An alternative solution to detect failure efficiently is to skip {\it uninteresting} functions (function {\tt f} in the example) before execution.
When the execution goes to the statement that depends on the skipped function (i.e., Line 9), it recovers the execution of function {\tt f} and merges the remaining states to avoid unsound results.
However, chopped execution relies heavily on the predefined skipped functions, which require prior expert knowledge to obtain.
Furthermore, the recovery mechanism, which switches between recovering and normal execution, is memory-consuming (see more details in Section \ref{mcts::sec:evaluation:rq2:memory-usage}).

\smallskip
{\bf Our Solution: \ourSol}. Unlike existing solutions, \ourSol performs vulnerability-oriented path search.
The basic idea behind \ourSol is that, before execution, we obtain the type-unsafe pointer locations (e.g., Lines 9, 16, and 20), where the type-unsafe pointers approximate the existence of vulnerabilities (i.e., the {\tt abort} failure in this example).
Then, guided by the locations, \ourSol will give a higher reward to those states with unsafe pointers, navigating the exploration towards the path where the number of unsafe pointers is maximized, i.e.,
\tikz[baseline=(char.base)] \node[shape=circle,draw,fill=white!20,inner sep=1pt, minimum size=4mm, font=\footnotesize] (char) {1}; ->
\tikz[baseline=(char.base)] \node[shape=circle,draw,fill=orange!20,inner sep=1pt, minimum size=4mm, font=\footnotesize] (char) {3}; ->
\tikz[baseline=(char.base)] \node[shape=circle,draw,fill=orange!20,inner sep=1pt, minimum size=4mm, font=\footnotesize] (char) {6}; ->
\tikz[baseline=(char.base)] \node[shape=circle,draw,fill=orange!20,inner sep=1pt, minimum size=4mm, font=\footnotesize] (char) {13};.
As a result, \ourSol explores the state 
\tikz[baseline=(char.base)] \node[shape=circle,draw,fill=orange!20,inner sep=1pt, minimum size=4mm, font=\footnotesize] (char) {13};
that triggers the  {\tt abort} failure at the top-1 position.

It is worth noting that although the existing search strategies in {\it Direction 1 \& 2} can ultimately trigger the {\tt abort} failure in the tiny example, due to the large spaces to explore in more realistic path exploration over complex software systems, they are very likely to miss important vulnerabilities due to inefficiency. 
In contrast, \ourSol could effectively explore the most promising paths that are more likely to contain vulnerabilities in practice.

\section{Approach}  \label{mcts::sec:approach}

{\bf Overview.} The general procedure of \ourSol is to continuously select a promising state that is more likely to contain vulnerabilities. 
Technically, \ourSol first acquires a set of unsafe pointers by performing pointer type inference over the test program only once at compile-time. After that, guided by unsafe pointers, \ourSol incorporates a new search strategy (i.e., MCTS) to have an optimal
balance between the exploration of future states and the exploitation of past executed states at execution time.
The main technical contribution of \ourSol lies in a new symbolic execution engine that implements a new path exploration strategy, based on a variant of MCTS equipped with the unsafe pointer-guided node expansion and the customized simulation policy, aiming to detect new vulnerabilities in software systems.

\subsection{Acquisition of Type-Unsafe Pointers via Type Inference System}

This subsection first justifies why the type-unsafe pointer is a reasonable indicator to approximate unknown memory safety vulnerability, and then articulates how \ourSol collects them.

\subsubsection{Why Type-Unsafe Pointers?}

The main root cause of memory safety problems in C/C++ is due to sacrificing type safety for flexibility and performance in the early design choice in the 1970s \cite{ccured-popl2002}.
As mentioned in Section \ref{mcts::sec:back-memory-safety}, to ensure spatial memory safety, it is essential to track specific properties (e.g., size and types) of the memory area to which the pointer refers. 
A type inference system is a well-established static analysis technique that was used to keep spatial memory safe in the literature \cite{szekeres2013sok,elliott2018checked}.
For example, \ccured \cite{ccured-popl2002} keeps spatial memory safe by classifying pointer types based on the usage of the pointers, while {\it SAFE} pointers that are free of memory errors can be soundly determined at compile time. For others (i.e., {\it SEQ} or {\it DYN}), memory safety must be ensured at run-time, which requires the insertion of safety checks during execution. 

Since there are no unified indicators to approximate unknown memory safety vulnerabilities, we suggest leveraging pointer types (i.e., {\it SEQ} and {\it DYN}) that can not be statically verified to be free of memory errors as {\it unsafe} pointers and use them to approximate vulnerable behaviors. We assume that if a program path contains more unsafe pointers, the path is more likely to contain vulnerabilities (the results shown in Figure \ref{mcts::fig:correlation} also support our claim). 

\subsubsection{How Does \ourSol Acquire Type-Unsafe Pointers?}
We follow an existing type inference algorithm {\tt \ccured} \cite{dataGuard} to classify pointer types according to the following three rules:

\begin{enumerate}[leftmargin=2.5em,nosep]
    \item All pointers are classified as {\it SAFE} upon their declaration.
    \item {\it SAFE} pointers that are subsequently used in pointer arithmetic are re-classified as {\it SEQ}.
    \item {\it SAFE} or {\it SEQ} pointers that are interpreted with different types are re-classified as {\it DYN}.
\end{enumerate}

This design provides sound, conservative pointer classification, guaranteeing that no pointer labeled {\it SAFE} is actually used unsafely \cite{nesCheck,ccured-popl2002}. 
In other words, \ourSol may produce false positives by marking some {\it non-DYN} pointers as {\it DYN}, but crucially avoids false negatives—it never marks {\it DYN} pointers as {\it non-DYN}.

\begin{algorithm}[t] 
    \small
    \caption{Unsafe Pointer-guided MCTS State Selection} \label{mcts::alg:mcts-overall}
    \KwIn{unsafe pointer set {\it unsafeSet}, a current state  {\it cur\_state}} 
    \KwOut{an execution state to be executed next {\it n\_state}}
    \SetKwFunction{FMain}{MCTSSearch::selectState}
    \SetKwProg{Fn}{Function}{:}{}
    \Fn{\FMain{}}{
        ExecutionState {\it n\_state} \\
        \While {(! {\tt isTerminal}(cur\_state->node))}{ \label{mcts::alg:mcts-overall:3}
            \If {(! {\tt hasEligibleChildren}(cur\_state->node))}{ \label{mcts::alg:mcts-overall:4}
                 {\it n\_state} = {\tt doSelection}(cur\_state->node)\label{mcts::alg:mcts-overall:5} \\
                cur\_state = {\it n\_state} \\
            }
            \Else {
                {\small n\_state = {\tt doExpansion}(cur\_state->node)}\label{mcts::alg:mcts-overall:8} \\
                \If {({\tt isWorthSimu}(cur\_state->node))}{ \label{mcts::alg:mcts-overall:9}
                    {\small reward={\tt doSimulation}(n\_state->node,unsafeSet)}\label{mcts::alg:mcts-overall:10} \\
                    {\small {\tt doBackpropagation}(reward, n\_state->node)}\label{mcts::alg:mcts-overall:11} \\
                }
                break
            }
        }
        \Return {\it n\_state} \label{mcts::alg:mcts-overall:12}
    }
\end{algorithm}

\subsection{Type-unsafe Pointer-guided Monte Carlo Tree Search}

The goal of a search strategy in symbolic execution is to select an interesting state to execute next.
Algorithm \ref{mcts::alg:mcts-overall} shows the overall selection strategy designed in \ourSol.
It takes a set of unsafe pointers {\it unsafeSet} acquired from the previous step and the current execution state being executed {\it cur\_state}, and outputs the expected state to be executed next.
There are many tree search algorithms proposed in the literature, such as Greedy Best-First Search \cite{heusner2017understanding}, Bidirectional Search \cite{sturtevant2018brief}, Uniform Cost Search \cite{cicalese2016tree}, and MCTS \cite{mcts2012survey}.
We chose MCTS in this study because it excels at maintaining a good trade-off between exploring new states in the future and exploiting known states in the past. 
The overall workflow of Algorithm \ref{mcts::alg:mcts-overall} customizes the standard MCTS algorithm with a few key steps guided by the unsafe pointer set.
Before diving into the details of the algorithm, we want to clarify that we use the process tree (internal data structure supported in KLEE \cite{klee}) as the symbolic execution tree to be used for MCTS.
Since every node represents an execution state in the process tree, we will use the term node and state interchangeably in the following sections.

The algorithm starts by performing a {\it while-loop} to check if the current state {\it cur\_state} is a terminal state or not (Line \ref{mcts::alg:mcts-overall:3}). 
If not, it checks whether the current state has eligible (i.e., expandable tree node) children via function {\tt hasEligibleChildren} (Line \ref{mcts::alg:mcts-overall:4}). 
The result will be either going back to the {\it while-loop} after performing node selection via {\tt doSelection} (Algorithm \ref{mcts::alg:doSelection}) in the {\it if-true} branch (Line \ref{mcts::alg:mcts-overall:5}) or calling {\tt doExpansion} (Algorithm \ref{mcts::alg:doExpansion}) in the {\it if-false} branch (starting at Line \ref{mcts::alg:mcts-overall:8}).
After the node expansion, it simulates the expanded node. It gets a reward by invoking {\tt doSimulation} (Algorithm \ref{mcts::alg:doSimulation} at Line \ref{mcts::alg:mcts-overall:10}) if the function {\tt isWorthSimu} return {\it true} (Line \ref{mcts::alg:mcts-overall:9}). A node is worth simulating when it has never been simulated or does not reach the simulation limit to avoid an infinite loop. Later, the reward is backpropagated through {\tt doBackpropagation} function to all the parent nodes until the root. 
Finally, the expanded state is returned as normal (Line \ref{mcts::alg:mcts-overall:12}).
It is worth noting that some key steps, such as node expansion and simulation, take actions based on the unsafe pointer set as one of the function parameters, where the record of unsafe pointers is essential to guide the smart search process in MCTS. We will explain each step in the following.

\subsubsection{Tree Node Selection.} \label{mcts::approach:selection}

The goal of the tree node selection is to select a child node that is already in the search tree.
Algorithm \ref{mcts::alg:doSelection} shows the overall procedure of node selection. 
Given the input of a tree node, it checks whether both left and right nodes exist in the search tree. 
If both nodes are valid, it selects the best child by calling {\tt selectBestChild} (Line \ref{mcts::alg:doSelection:3}).
Otherwise, only the valid left or right node will be selected (Lines \ref{mcts::alg:doSelection:4}-\ref{mcts::alg:doSelection:7}).

Inside function {\tt selectBestChild}, the child node is selected based on the highest UCT (i.e., Upper Confidence bounds applied to Trees) value.
UCT is a widely recognized algorithm that addresses a significant limitation of MCTS \cite{kocsis2006bandit,mcts2012survey,liu2020legion}, where the MCTS may incorrectly favor a suboptimal move that has a limited number of forced refutations.
The UCT formula used to balance the exploitation and exploration is defined as ${\small  UCT(s, s') = \frac{R(s')}{V(s')}+C\sqrt{\frac{2\ln V(s)}{V(s')}}}$ (details can be found in Section 3.3 in \cite{mcts2012survey}),
where $s$ is the current state, $s'$ is the child state of state $s$ being selected, $V(s)$ indicates how many times the state has been visited, and $R(s)$ is the cumulative reward of all the simulations that have passed through this state.
$C$ is a constant that controls the exploration degree. 

\begin{algorithm}[h] 
    \small
    \caption{Procedure of Tree Node Selection in \ourSol} \label{mcts::alg:doSelection}
    \KwIn{a tree node {\it node} (with attributes such as {\it isInTree})} 
    \KwOut{a leaf node of {\it node} to be selected {\it selected\_node}}
    \SetKwFunction{FMain}{\texttt{doSelection}}
    \SetKwProg{Fn}{Function}{:}{}
    \Fn{\FMain{node}}{
        \If {node->left->isInTree \&\& node->right->isInTree}{
            selected\_node = {\tt selectBestChild}(node)\label{mcts::alg:doSelection:3}
        }
        \If {node->left->isInTree \&\& ! node->right->isInTree}{\label{mcts::alg:doSelection:4}
            selected\_node = node->left
        }
        \If {! node->left->isInTree \&\& node->right->isInTree}{
            selected\_node = node->right 
        }\label{mcts::alg:doSelection:7}
        \Return selected\_node \label{mcts::alg:doSelection:8}
    }
    \SetKwFunction{FMain}{\texttt{selectBestChild}}
    \SetKwProg{Fn}{Function}{:}{}
    \Fn{\FMain{node}}{
        \tcc{return the node with the highest {\it UCT} value}
    }           
\end{algorithm}

\rev{When a node is selected (Line \ref{mcts::alg:doSelection:8}), we check whether the node has any successors that have not yet been added to the search tree. If such a child exists, {\tt doExpansion} is performed; otherwise, if the node is fully expanded, the algorithm returns to the while-loop to continue selection from the best child node of the successors.}

\subsubsection{Tree Node Expansion.} \label{mcts::approach:expansion}

The expansion aims to expand a child that is {\it not} in the search tree yet.

Algorithm \ref{mcts::alg:doExpansion} presents the overall procedure of tree node expansion. Given an input node, different from node selection, it checks whether both left and right nodes do not exist in the search tree. 
If neither node is in the search tree, it expands the best child by calling {\tt expandBestChild} (Line \ref{mcts::alg:doExpansion:3}).
Otherwise, only the left or right node will be augmented for expansion (Lines \ref{mcts::alg:doExpansion:4}-\ref{mcts::alg:doExpansion:7}).
Inside the function {\tt expandBestChild}, the node is expanded based on the highest expansion score ({\it ExpScore}) among the two branches. 

To obtain the score of two branches after an execution state is forked, three steps are performed: 
(1) collecting unique program elements, i.e., basic blocks, that are \emph{unique} to each state;
(2) analyzing the collected basic blocks and finding a way to weigh which state is more vulnerable;
and (3) scoring the two branches based on the weighting measurement.

\begin{algorithm}[h]
    \small
    \caption{Procedure of Tree Node Expansion in \ourSol} \label{mcts::alg:doExpansion}
    \KwIn{a tree node {\it node}, (global) unsafe pointer set {\it unsafeSet}} 
    \KwOut{a leaf node of {\it node} to be expanded {\it expanded\_node}}
    \SetKwFunction{FMain}{\texttt{doExpansion}}
    \SetKwProg{Fn}{Function}{:}{}
    \Fn{\FMain{node}}{
        \If {! node->left->isInTree \&\& ! node->right->isInTree}{
            expand\_node = {\tt expandBestChild}(node, unsafeSet)\label{mcts::alg:doExpansion:3}
        }
        \If {node->left->isInTree \&\& ! node->right->isInTree}{\label{mcts::alg:doExpansion:4}
            expanded\_node = node->right
        }
        \If {! node->left->isInTree \&\& node->right->isInTree}{
            expanded\_node = node->left 
        }\label{mcts::alg:doExpansion:7}
        \Return expanded\_node \label{mcts::alg:doExpansion:8}
    }
    \SetKwFunction{FMain}{\texttt{expandBestChild}}
    \SetKwProg{Fn}{Function}{:}{}
    \Fn{\FMain{node, unsafeSet}}{ \label{mcts::alg:doExpansion:9}
        \tcc{return the node with the highest {\it ExpScore} value}
    }         
\end{algorithm}

\begin{figure}[t]
	\centering
	\includegraphics[width=0.5\linewidth]{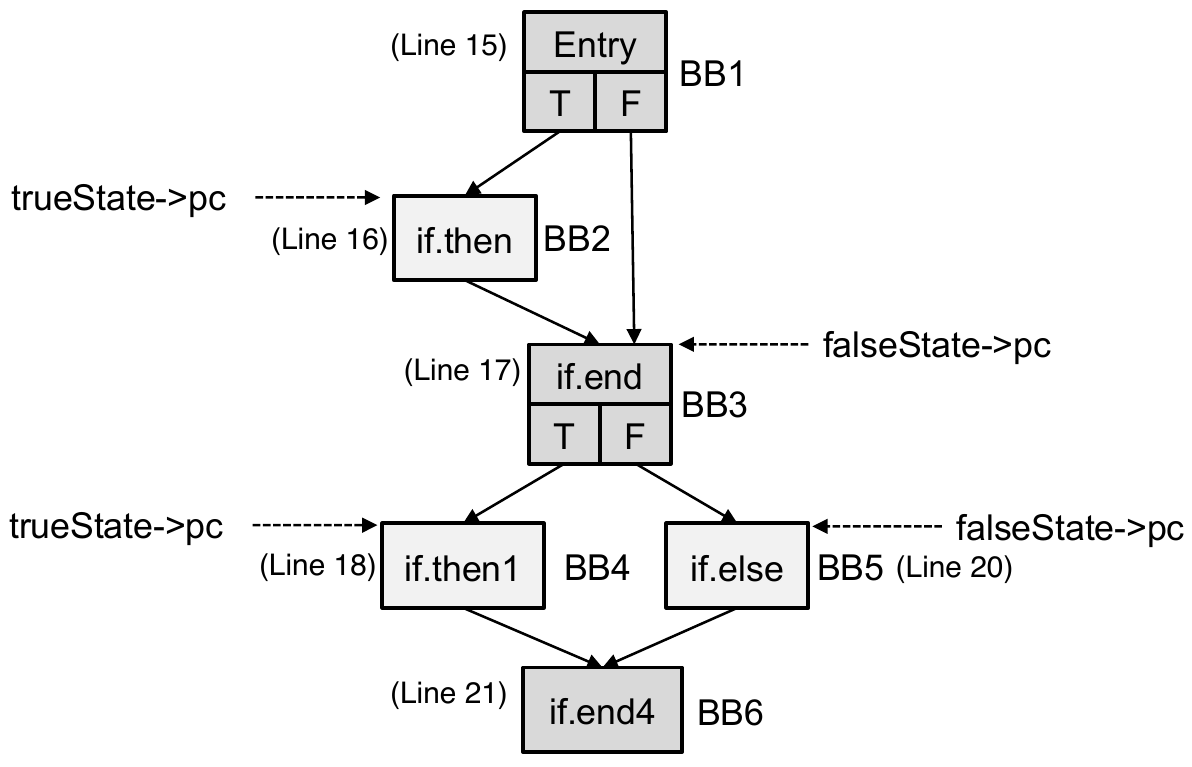}
	\caption{CFG of function {\tt f} shown in Figure \ref{mcts::fig:motivating-example}(a) illustrating how to get the score of true/false states}
	\label{mcts::fig:getExpScore}
\end{figure}

To accomplish the first step, we use the dominance relationships of graph theory \cite{lowry1969object} on the Control Flow Graph (CFG) of a function to collect the unique basic blocks for each state. 
The idea is to find the {\it post-dominator} of the two new forked states and to collect the basic blocks in between each of the states and their post-dominator; such basic blocks indicate how the two states may induce different executions.
A node $A$ on a graph is said to {\it dominate} another node $B$ if every path from the graph's entry point (start node) to $B$ must go through $A$.
Conversely, a node $A$ is a {\it post-dominator} of another node $B$ if every path from $B$ to the exit point (or end node) of the graph must pass through $A$.
Taking the function {\tt f} in the example shown in Figure \ref{mcts::fig:motivating-example} as an illustration, the CFG of the function is shown in Figure \ref{mcts::fig:getExpScore}.
Starting from the entry block ({\tt BB1}), the execution will go to Line 15 and the engine will fork two states ({\tt TrueState} and {\tt FalseState}), where the {\it pc} (points to the next execution instruction) {\tt TrueState} points to the basic block ({\tt BB2}) starting at Line 16. 
The {\it pc} in {\tt FalseState} points to the basic block ({\tt BB3}) after the {\it if-branch} at Line 15. The unique basic block for {\tt TrueState} is {\tt BB2}, as both states will go through the post-dominator block {\tt BB3}.
For the following forking at Line 17 in Figure \ref{mcts::fig:motivating-example}(a), again, the unique basic block for two states is {\tt BB4} and {\tt BB5}, respectively.

Then, following the intuition that more unsafe pointers in a path imply more vulnerabilities, we use the number of unsafe pointers in the basic blocks to quantify the expansion scores of the states:
{\it ExpScore(s) =} $N_{unsafe}(s)$,
where {\it s} is a state being scored and $N_{unsafe}(s)$ represents the number of unsafe pointers 
\cc{in the basic blocks collected as above for $s$}.

\subsubsection{Tree Node Simulation} \label{mcts::approach:simulation}

Simulation is a crucial component of the MCTS algorithm as it enables the evaluation of non-terminal nodes by performing random 
\cc{playouts (i.e., simulation runs)}
to estimate the potential outcomes (i.e., rewards) of different actions.
This process helps balance exploration and exploitation by providing statistical sampling to approximate the reward of execution states, which guides the algorithm in making promising decisions. 

\cc{There can be many options to perform each simulation run. Intuitively, for our symbolic execution, we could consider the following three options (OPs)}
to get the reward of a node.

\begin{itemize}[leftmargin=1em,nosep]
    \item {\it OP1} statically analyzes a single path following the state based on control-flow graphs.
    \item {\it OP2} symbolically runs a single feasible path following the state until a fixed number of executed instructions are reached.
    \item {\it OP3} performs symbolic execution along a single feasible path until the corresponding execution state is terminated.
\end{itemize}

Note that a high-quality simulation policy should be efficient, have few false positives, and have a certain degree of randomness \cite{mcts2012survey,liu2020legion}. 
As shown in the literature \cite{mcts2012survey}, a certain degree of randomness is advantageous as it is simple and requires no domain knowledge, which will most likely cover diverse areas of a search space.

Concerning the above criteria, {\it OP1} might be good in terms of randomness, but efficiency and false positives may be an issue: 
Computing CFGs may be costly, and paths in static CFGs may be infeasible.
For {\it OP2}, it may comply with the criteria, but it might be difficult to decide what is the preferable number of instructions to use when evaluating the reward of a node across different runs.
{\it OP3} could be better as it terminates based on the behavior of program execution.
Technically, the simulation execution is essentially a lightweight depth-first execution of a single random feasible path.
When a node is expanded, a new node is forked out for simulation. The new node's path constraints remain, and the simulation is started by setting the node attribute ``{\it inhibitForking=true}'' to prevent further forking. Then, whenever a branch condition is encountered during simulation, a feasible random branch is chosen until the execution terminates.
We empirically evaluate different fixed instruction numbers for {\it OP2}, and the results show that they are inferior compared with {\it OP3} (see more details in Section \ref{mcts::sec:evaluation:rq3}).

Algorithm \ref{mcts::alg:doSimulation} shows the simplified procedure of the simulation process designed in \ourSol. 
It takes a tree node to be expanded and the unsafe pointer set as input, and yields a reward after simulation. Inside the algorithm, a vector of basic blocks {\it simulatedBB} is first initiated (Line \ref{mcts::alg:doSimulation:2}). 
The vector stores all basic blocks executed during simulation via the function {\tt executor.sim\_run}.
Then, two metrics (the number of unsafe pointers {\it nu\_unsafe} and memory errors {\it nu\_error}) are calculated (Line \ref{mcts::alg:doSimulation:4}) or updated (Line \ref{mcts::alg:doSimulation:5}), which will be used to get the reward of this simulation run (Line \ref{mcts::alg:doSimulation:6}).
In this study, \ourSol is novel in utilizing the random and lightweight simulation runs to identify the likely vulnerable areas, and then utilizing MCTS search to guide the symbolic execution capable of state forking to move toward those areas.

\smallskip
\begin{algorithm}[h] 
    \small
    \caption{Procedure of Tree Node Simulation in \ourSol} \label{mcts::alg:doSimulation}
    \KwIn{a tree node {\it node}, the unsafe pointer set {\it unsafeSet}} 
    \KwOut{reward of the simulation {\it reward}}
    \SetKwFunction{FMain}{\texttt{doSimulation}}
    \SetKwProg{Fn}{Function}{:}{}
    \Fn{\FMain{node, unsafeSet}}{
        vector<BasicBlock*> simulatedBB \label{mcts::alg:doSimulation:2} \\ 
        simulatedBB = {\tt executor.sim\_run}(node) \label{mcts::alg:doSimulation:3} \\
        nu\_unsafe = {\tt getNumUnsafePt}(simulatedBB, unsafeSet)\label{mcts::alg:doSimulation:4} \\
        nu\_error = {\tt executor.numOfMemError}() \label{mcts::alg:doSimulation:5} \\
        reward = {\tt getReward}(nu\_unsafe, nu\_error) \label{mcts::alg:doSimulation:6} \\
        \Return reward \label{mcts::alg:doSimulation:7}
    }
\end{algorithm}

When a simulation execution terminates, the reward function is used to calculate the reward for this node. We define the reward function as follows: 
$F_{reward}(s) = 0.5 * N_{unsafe}(s) + 0.5 * N_{error}(s)$,
where the $N_{unsafe}(s)$ denotes the number of unsafe pointers covered during the simulated run and $N_{error}(s)$ is the number of memory errors detected throughout the run. 
We include the detected number of memory errors in the reward function based on a common observation that a root cause of one memory error may lead to various error symptoms in many code regions \cite{fix2bugs} and code regions related to a buggy code region likely contain bugs too, so we include this number in the reward expecting more memory errors can be found.

\smallskip
{\it Simulation Optimization.} 
When simulating a node in a for-loop statement, the default simulation process repeatedly simulates the same node. 
Such a process can be inefficient and waste a lot of time (see more evaluation results in Section \ref{mcts::sec:evaluation:rq3}).
Therefore, we design a new simulation optimization strategy in \ourSol,
to reduce the simulation of {\it unimportant} states that bring little new rewards due to the existence of loops.
A straightforward way to avoid an unlimited simulation on a loop is to set the loop bound when doing a simulation. However, this may still lead to unsound results \cite{survey-se2018}. 
\cc{Our insight is that instead of setting a fixed loop bound for the simulation, we consider a degree of increment based on the rewards of the previous simulation on demand. 
Therefore, our solution is that when simulating a node in a loop, we record its reward after each simulation run of the loop. If it cannot get a higher reward after a limited number of iterations (we provide an extra option ``{\it --optimization-degree}'' to allow users flexibly control the number of loop iterations), it will not repeat the simulation of that node.
}

\subsubsection{Backpropagation}
Backpropagation is the final phase in MCTS-guided sampling, during which the total reward and visit count for each state are iteratively updated. For each state, the reward is adjusted based on the simulation outcome, and the visit count is incremented, with updates propagating from the leaf node to the root (performed at Line 11 in Algorithm \ref{mcts::alg:mcts-overall}).

\subsection{Implementation of \ourSol}  \label{mcts::sec:implementation}

We implemented \ourSol on top of KLEE (v3.0). 
Following the instructions on the webpage, users can set up and run \ourSol to find potential vulnerabilities in the test programs automatically.

Both the implementation of the type inference system and the MCTS algorithm are written in the C++ programming language.
For the type inference system used in \ourSol, we forked {\tt \ccured} (built on top of a static analysis tool SVF \cite{sui2016svf}) from a previous work \cite{dataGuard}. 
We added a new search strategy {\tt MCTSSearcher} in the {\tt Search} class in KLEE's implementation to support the vulnerability-oriented path exploration.
In general, we implemented our own {\tt selectState} and {\tt update} to maintain execution states.
The functionality of the MCTS algorithm is implemented in the {\tt selectState} function, where four major functions (i.e., {\tt doSelection}, {\tt doExpansion}, {\tt doSimulation}, and {\tt doBackpropagation}) mentioned in the Algorithm \ref{mcts::alg:mcts-overall} are supported.
We also modified the {\tt executeInstruction} and {\tt run} functions in {\tt Executor.cpp} to support the checking of unsafe pointers during execution and the storing of the executed unsafe pointers into a file (``{\it unsafe-pt.txt}'') to help users analyze the program further.

For setting the bias parameter {\it C} in Equation in Section \ref{mcts::approach:selection} and the simulation optimization option ``{\it -- optimization-degree}'', we set the value of $\sqrt{2}$ and 700 for them after evaluating their impacts, respectively (more detailed results and discussion are presented in Section \ref{mcts::sec:discussion:impact}).

\section{Evaluation}  \label{mcts::sec:evaluation}

Extensive experiments are conducted to evaluate the effectiveness of the proposed \ourSol from various perspectives. More specifically, we consider the following \cc{four} research questions (RQs).

\begin{itemize}[leftmargin=1em,nosep]
    \item {\bf RQ1}: Can \ourSol cover more unsafe pointer?
    \item {\bf RQ2}: Can \ourSol effectively detect known vulnerabilities?
    \item {\bf RQ3}: Can each major component contribute to \ourSol?
    \item {\bf RQ4}: Can \ourSol detect new (i.e., previously unknown) vulnerability?
\end{itemize}

Among these RQs, RQ1 evaluates \ourSol's capabilities in terms of unsafe pointer coverage and memory error detection compared with representative path exploration strategies in KLEE and CBC.
RQ2 aims to investigate the vulnerability detection capabilities in terms of time efficiency and memory usage compared with other approaches over known CVE vulnerabilities.
RQ3 concentrates on the contribution of each component of \ourSol.
RQ4 demonstrates the practical vulnerability detection capability of \ourSol.

All experiments conducted in this study ran on a Linux PC with Intel(R) Xeon(R) W-2133 CPU @ 3.60GHz x 12 processors and 64GB RAM running Ubuntu 18.04 operating system.

\subsection{RQ1: Unsafe-pointer Covering Capability} \label{mcts::sec:evaluation:rq1}

{\bf Benchmarks.} We use the well-known GNU {\tt Coreutils} (v9.5) in this RQ1, following many existing works \cite{klee,trabish2018chopped,tu2022fastklee,steering-se,ccs21-learch}. 
The utilities include the basic file, shell, and text manipulation tools of the operating system.
We selected 75 utilities for this study and excluded some utilities that: (1) cause nondeterministic behaviors (e.g., \texttt{kill}, \texttt{ptx}, and \texttt{yes}) and (2) exit early due to the unsupported assembly code or the call for external functions. 

For the selected utilities, we measure the size of executable lines of code (ELOC) by counting the total number of executable lines in the final executable after global optimization.
The distribution of ELOC ranges from 800-8,000, which could comprehensively evaluate the effectiveness of \ourSol on test programs of various lengths.

\smallskip
{\bf Approaches and Evaluation Metrics for Comparison.} 
We select six representative path exploration strategies as follows, including three commonly evaluated in the literature \cite{trabish2020relocatable,steering-se,tu2022fastklee,symloc} and three coverage-guided heuristics:
\begin{itemize}[leftmargin=1em,nosep]
    \item Breadth first search ({\tt bfs}) and depth first search ({\tt dfs}).
    \item Random ({\tt random-state}) randomly selects a state to explore.
    \item Code coverage-guided ({\tt nurs:covnew}) selects a state that has a better chance to cover new code.
    \item Instruction coverage-guided ({\tt nurs:md2u}) prefers a state with minimum distance to an uncovered instruction, while ({\tt nurs:icnt}) picks a state trying to maximize instruction count.
    \item CBC \cite{yi2024compatible} proposes compatible branch coverage-driven path exploration for symbolic execution.
    \item \rev{\cgs \cite{cgs-icse24} targets on exploring concrete branches that are neglected during path exploration.}
    \item \rev{\featmaker \cite{featMaker-fse24} proposes a feature-driven path exploration strategy to improve code coverage.}
    \item \rev{\empc \cite{empc-sp25} utilizes path cover information for new path prioritization to boost path exploration.}
\end{itemize} 

\rev{Note that the default implementation of KLEE, CBC, \cgs, \featmaker, and \empc does not support the recording of unsafe pointer coverage, and we have modified their source code to enable this functionality for a fair comparison.}

We use the following two metrics to assess the effectiveness of different path search strategies:
\begin{itemize}[leftmargin=1em,nosep]
\item (1) {\it The number of unsafe pointers covered} compares the pointer covering capabilities of various approaches.
\item (2) {\it The number of memory errors detected} compares memory error detection capabilities of various approaches.
\end{itemize}

\smallskip
{\bf Running Setting.} Following existing studies \cite{klee,tu2022fastklee,steering-se}, we set a running time of one hour for each setting. 
For random searches, we ran them five times and reported the median results. 

\begin{table}[t]
    \small
    \centering 
    \caption{\rev{Results compared with prior search strategies}} 
    \vspace{-1em}
    \setlength{\tabcolsep}{8pt} 
    \begin{tabular}{lrrrr|rr}
        \hline
        \multirow{2}{*}{\textbf{Strategies}} & \multicolumn{4}{c|}{{\bf Unsafe Pointers}} & \multicolumn{2}{c}{{\bf Memory Errors}}  
         \\
        \cline{2-7}
        & \textit{Total} & \textit{Imp$_{total}$} & \textit{Total$_{error}$} & \textit{Imp$_{error}$} & \textit{Total} & \textit{Imp$_{total}$}  \\
        \hline
        {\tt bfs} & 11610 & 11.01\% & 2361 & 14.95\% & 40 & 10.00\%  \\
        {\tt dfs} & 9626 & 33.89\%  & 1603 & 69.31\% & 32 & 37.50\%  \\
        {\tt random} & 7609 & 69.38\%  & 1825 & 48.71\% & 34 & 29.41\%  \\
        {\tt nurs:covnew} & 10232  & 25.96\% & 2685 & 3.39\% & 39& 12.82\%  \\
        {\tt nurs:md2u} & 9417 & 36.86\% & 1980 & 37.07\%  & 39 & 12.82\%  \\
        {\tt nurs:icnt} & 6782 & 90.03\% & 1635 & 66.00\%   & 35 & 25.71\% \\
        \rev{\cgs} \cite{cgs-icse24}  & \rev{8406} & \rev{53.53\%} & \rev{1296} & \rev{109.41\%} & \rev{25} & \rev{76.00\%}  \\
        \rev{\featmaker} \cite{featMaker-fse24}  & \rev{8185} & \rev{57.46\%} & \rev{1531} & \rev{77.27\%} & \rev{34} & \rev{29.41\%} \\
        \rev{\empc} \cite{empc-sp25}  & \rev{6098 (7145)$^*$} & \rev{17.17\%} & \rev{2096 (2289)$^*$} & \rev{9.21\%} & \rev{35 (38)$^*$} & \rev{8.57\%}  \\
        CBC \cite{yi2024compatible}  & 2194 (3890)$^*$ & 77.30\% & 597 (1168)$^*$ & 95.64\% & 21 (33)$^*$ & 57.14\%  \\
        \hline
        \ourSol & 12888  & - & 2714 & - & 44 & -  \\
        \hline
    \end{tabular}
    \begin{tablenotes}
        \footnotesize
        \item* CBC \cite{yi2024compatible} can only successfully run 50 out of 75 utilities, so when comparing to CBC, we only run \ourSol over the same 50 utilities as CBC, and the numbers in the parentheses are the numbers of unsafe pointers/memory errors covered/detected by \ourSol for the 50 utilities only.  
      \end{tablenotes}
    \label{mcts::tab:rq1}
\end{table}

\smallskip
{\bf Results.} 
Table \ref{mcts::tab:rq1} shows the overall results, where the numbers of two metrics and the improvement achieved by \ourSol are recorded.
In the table, the column \textit{Total} represents the total number of unsafe pointers covered or memory errors detected by each approach, while \textit{Imp} indicates the improvement (in percentage) achieved by \ourSol compared to each approach.
\textit{Total$_{error}$} represents the total number of memory errors detected by each approach, while \textit{Imp$_{error}$} indicates the improvement (in percentage) achieved by \ourSol compared to each approach.
From the table, it is evident that \ourSol performs significantly better than existing search strategies in terms of both the number of covered unsafe pointers and detected memory errors.
In particular, \ourSol could cover 90.03\% more unsafe points compared to {\tt nurs:icnt} and detect 57.14\% more memory errors compared to CBC, demonstrating the superior performance of \ourSol.
\rev{Compared to \cgs, \featmaker, and \empc, \ourSol also outperforms them in both metrics.
Specifically, \ourSol covers 53.53\%, 57.46\%, and 17.17\% more unsafe pointers in total and detects 109.41\%, 77.27\%, and 9.21\% more memory errors than \cgs, \featmaker, and \empc, respectively.
}
Note that following a prior work \cite{steering-se}, we did not submit these reports to developers, as we target evaluating detection accuracy, not responsible disclosure. 
This choice is justified by the fact that Vital builds on KLEE—a tool widely adopted in research and industry and known to produce true positives—so we follow KLEE's assumption that all reported errors (all out-of-bounds issues) are true positives.

\begin{figure*}[t]
	\centering
	\subfigure[\ourSol vs {\tt bfs}]{
	\begin{minipage}[t]{0.18\linewidth}
	\centering
	\includegraphics[width=2cm]{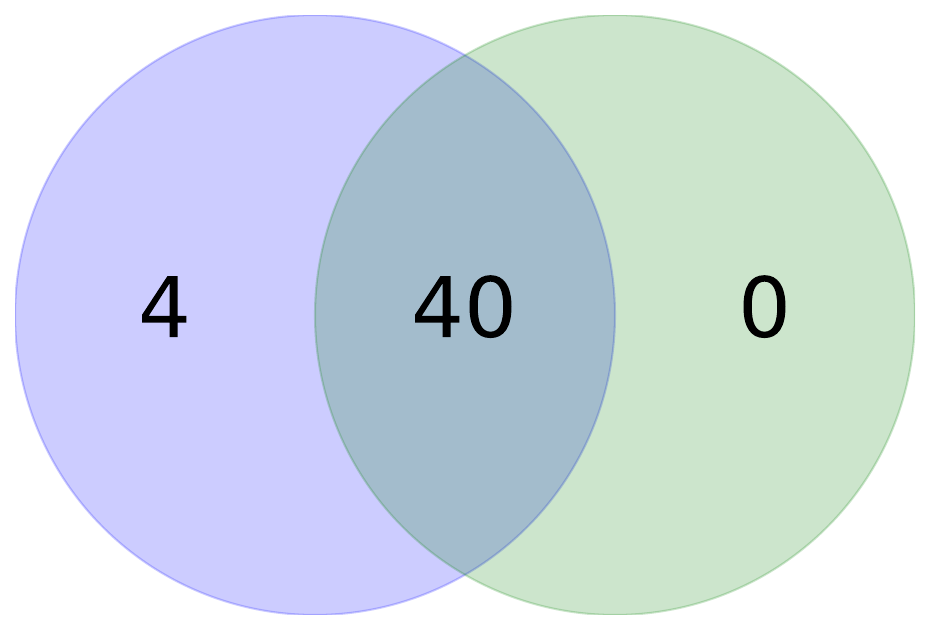}
	\vspace{-1em}
	\label{mcts::fig:error-bfs}
	\end{minipage}
	}
	\subfigure[\ourSol vs {\tt dfs}]{	
	\begin{minipage}[t]{0.18\linewidth}
	\centering
	\includegraphics[width=2cm]{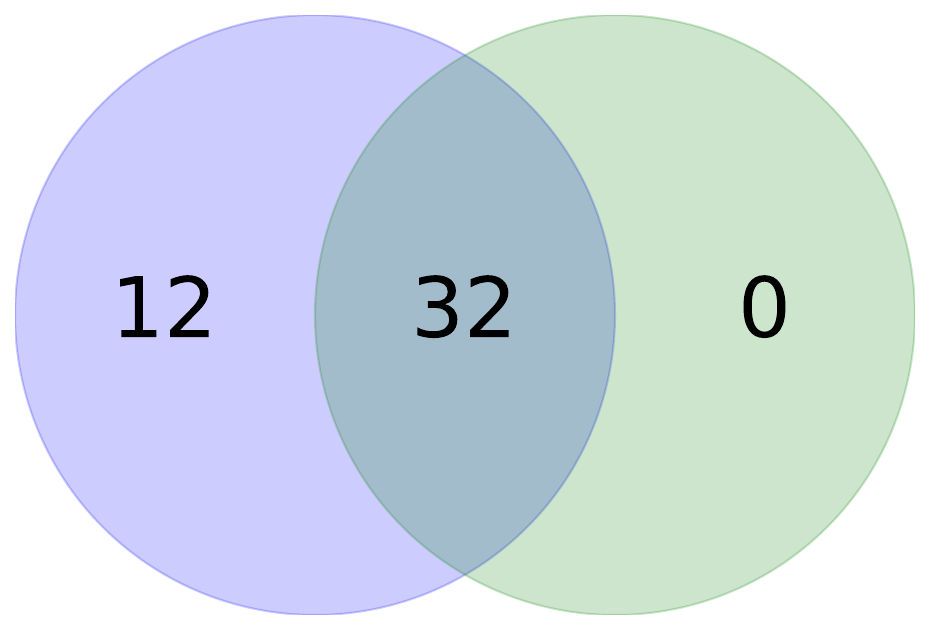}
	\vspace{-1em}
	\label{mcts::fig:error-dfs}
	\end{minipage}	
	}
    \subfigure[\ourSol vs {\tt random}]{	
	\begin{minipage}[t]{0.18\linewidth}
	\centering
	\includegraphics[width=2cm]{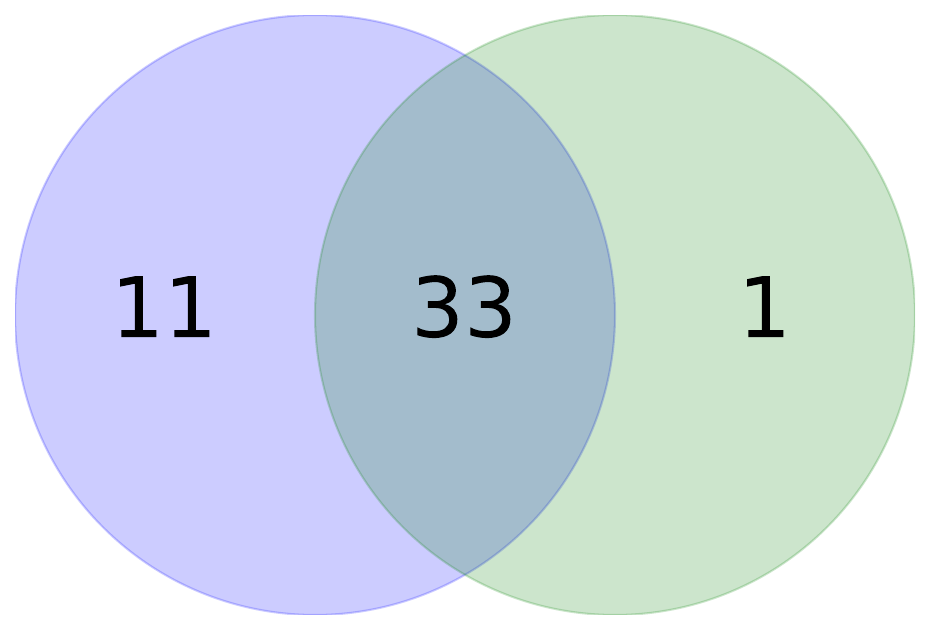}
	\vspace{-1em}
	\label{mcts::fig:error-random}
	\end{minipage}	
	}
    \subfigure[\ourSol vs {\tt covnew}]{	
	\begin{minipage}[t]{0.18\linewidth}
	\centering
	\includegraphics[width=2cm]{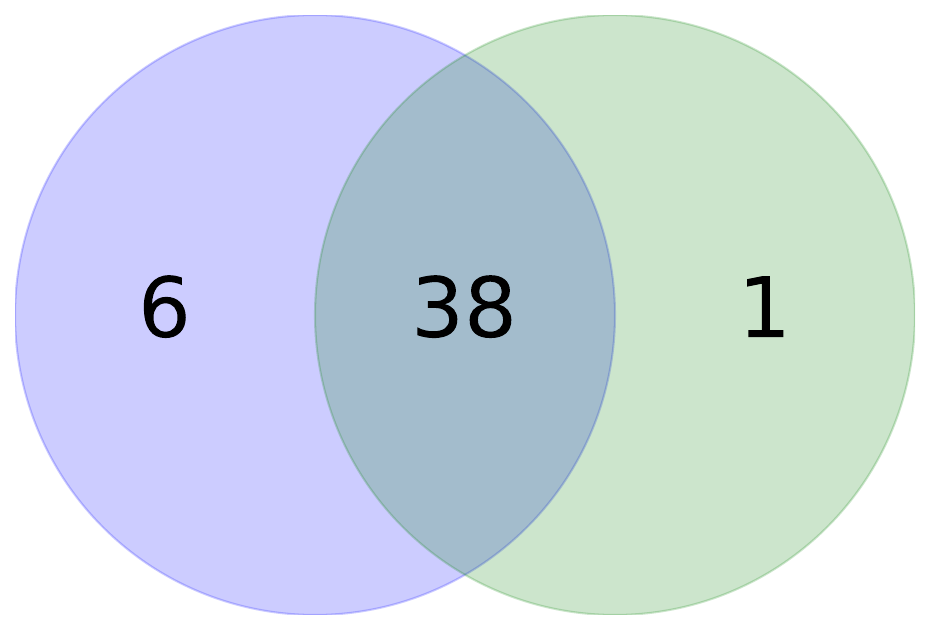}
	\vspace{-1em}
	\label{mcts::fig:error-covnew}
	\end{minipage}	
	}
    \subfigure[\ourSol vs {\tt md2u}]{	
	\begin{minipage}[t]{0.18\linewidth}
	\centering
	\includegraphics[width=2cm]{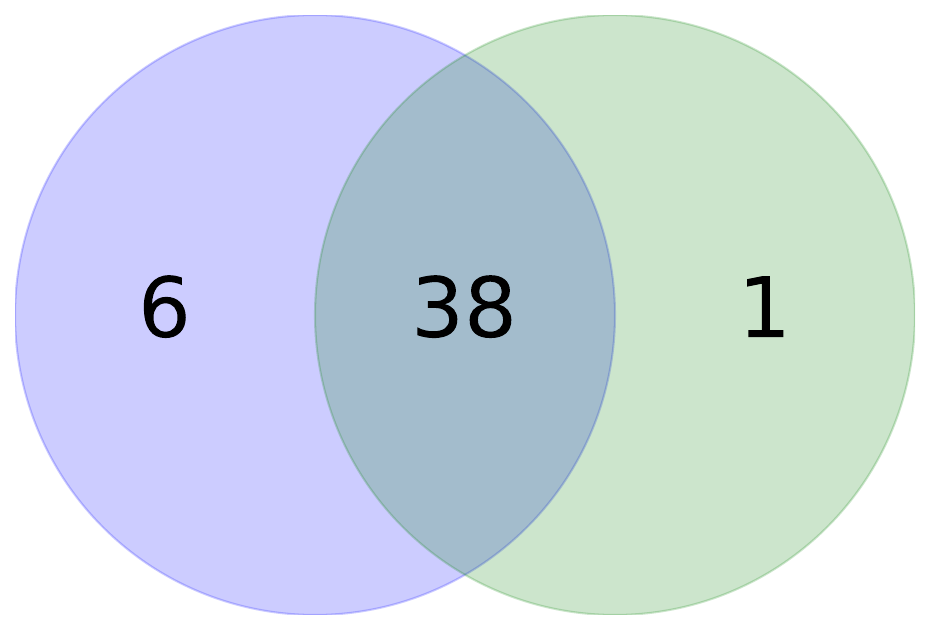}
	\vspace{-1em}
	\label{mcts::fig:error-md2u}
	\end{minipage}	
	}
    \subfigure[\ourSol vs {\tt icnt}]{	
	\begin{minipage}[t]{0.18\linewidth}
	\centering
	\includegraphics[width=2cm]{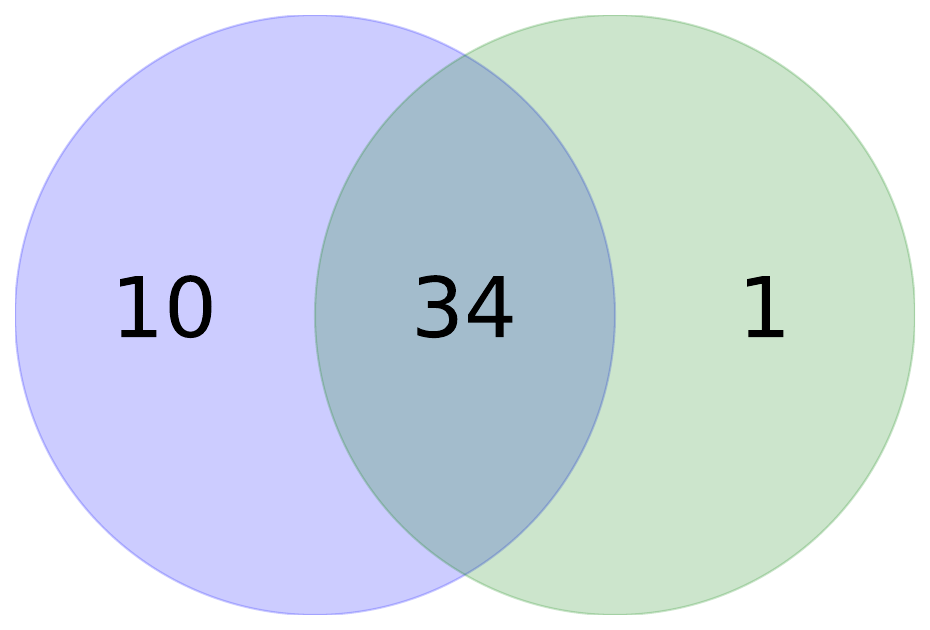}
	\vspace{-1em}
	\label{mcts::fig:error-icnt}
	\end{minipage}	
	}
    \subfigure[\ourSol vs CBC \cite{yi2024compatible}]{	
	\begin{minipage}[t]{0.18\linewidth}
	\centering
	\includegraphics[width=2cm]{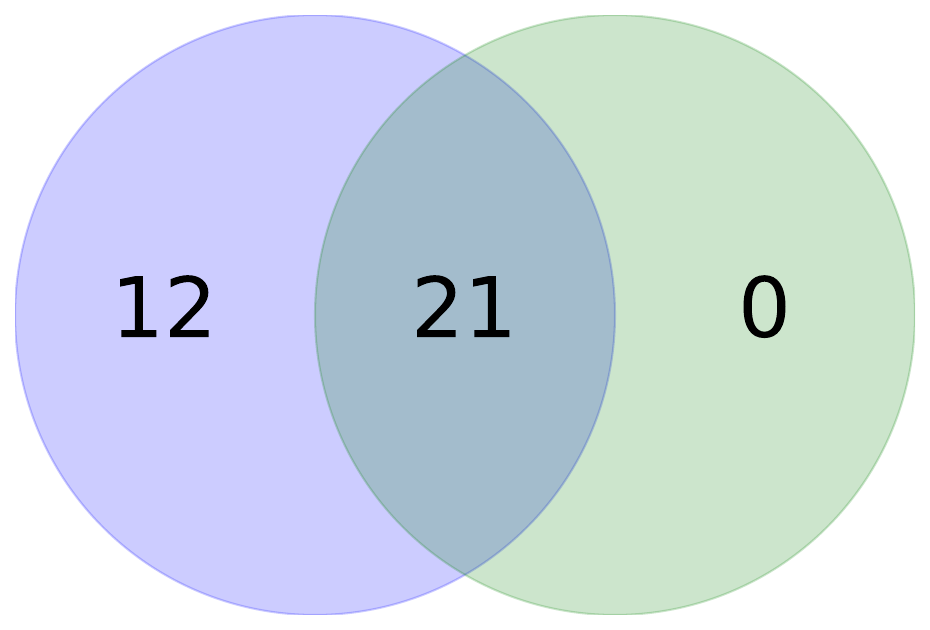}
	\vspace{-1em}
	\label{mcts::fig:error-icnt}
	\end{minipage}	
	}
    \subfigure[\rev{\ourSol vs \cgs \cite{cgs-icse24}}]{	
	\begin{minipage}[t]{0.18\linewidth}
	\centering
	\includegraphics[width=2cm]{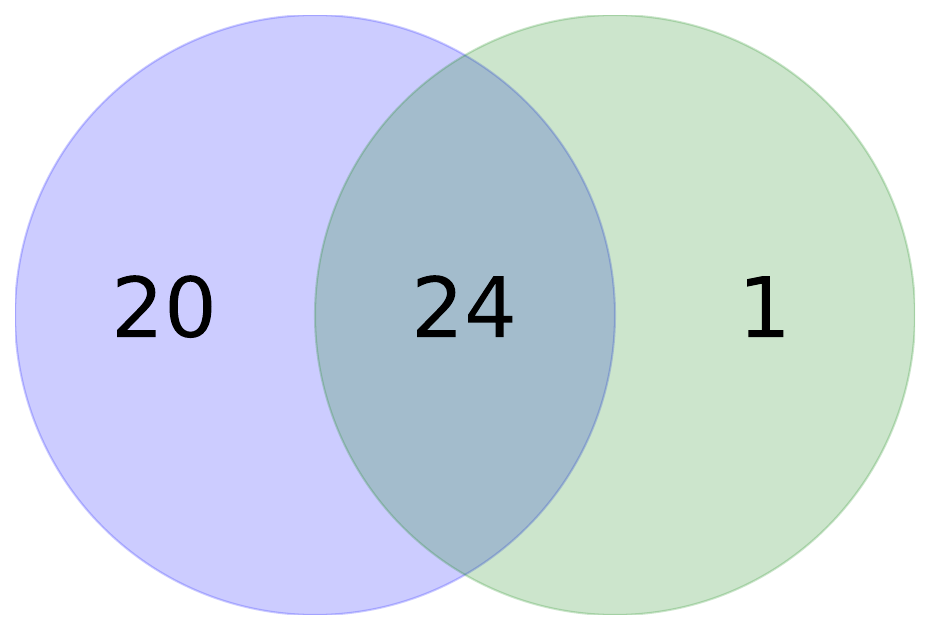}
	\vspace{-1em}
	\label{mcts::fig:error-icnt}
	\end{minipage}	
	}
    \subfigure[\rev{\ourSol vs  \textsc{Fea.} \cite{featMaker-fse24}}]{	
	\begin{minipage}[t]{0.18\linewidth}
	\centering
	\includegraphics[width=2cm]{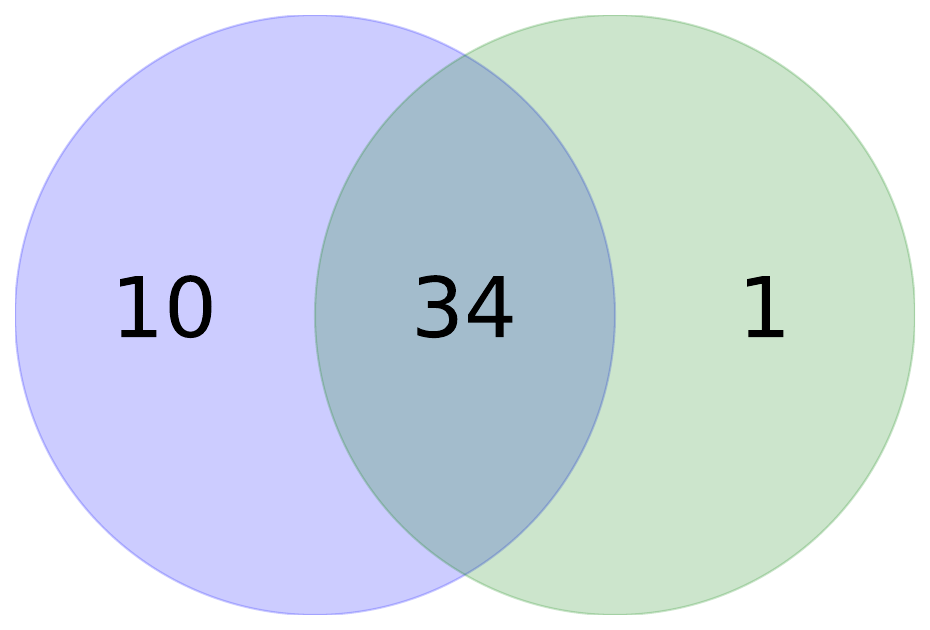}
	\vspace{-1em}
	\label{mcts::fig:error-icnt}
	\end{minipage}	
	}
    \subfigure[\rev{\ourSol vs \empc \cite{empc-sp25}}]{	
	\begin{minipage}[t]{0.18\linewidth}
	\centering
	\includegraphics[width=2cm]{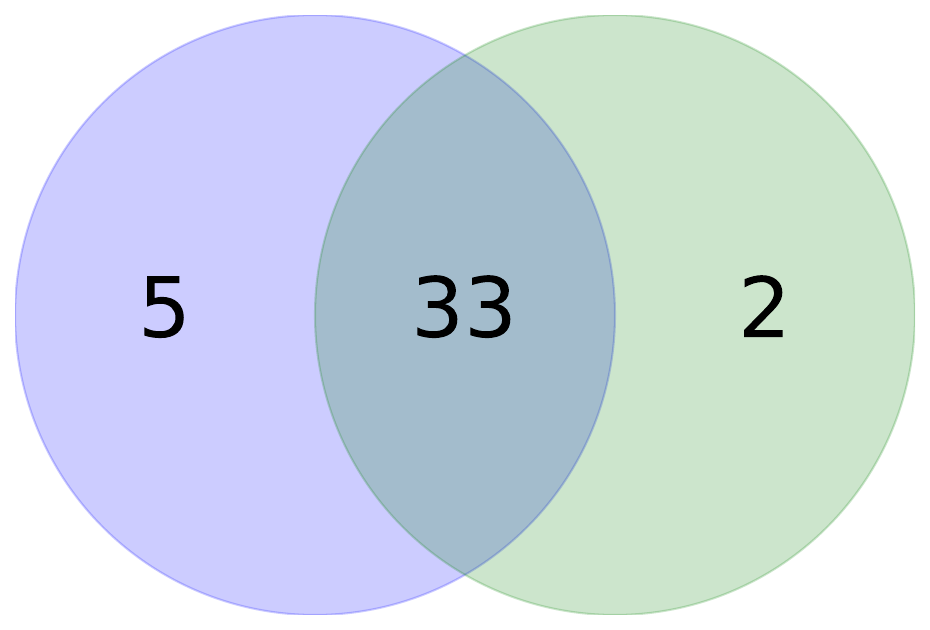}
	\vspace{-1em}
	\label{mcts::fig:error-icnt}
	\end{minipage}	
	}
	\vspace{-1em}
	\caption{\rev{Number of detected {\it unique} memory errors by \ourSol compared to other search strategies}}
	\label{mcts::fig:unique-memory-errors}
	\end{figure*}

We also investigate the {\it unique} memory error detection capability of \ourSol, and the Venn Figure~\ref{mcts::fig:unique-memory-errors} presents the overall results. 
From the figure, we can observe that \ourSol could detect larger numbers (ranging from 4 compared to {\tt bfs} and 20 to \cgs \cite{cgs-icse24}) of unique memory errors.
\rev{In particular, compared to the latest searching strategies in the last four Venn plots, \ourSol covers almost all memory errors detected by the baselines while contributing many additional ones. Compared to CBC, \ourSol contributes 12 unique detections and overlaps on 21, with CBC adding none. Compared to \cgs, \ourSol adds 20 unique overlaps on 24, and \cgs contributes only 1 unique case. Against \featmaker, \ourSol adds 10 unique with 34 overlaps and just 1 unique for \featmaker. Against \empc, \ourSol contributes 5 unique with 33 overlaps, and EMPC adds 2 unique. Overall, \ourSol captures 95–100\% of the combined detections in each pair and consistently provides the largest set of exclusive findings, indicating a better memory error detection capability.}
\ourSol misses only a few memory errors detected by other strategies (up to 2 when compared to \empc), this is mainly because the MCTS algorithm may be biased toward more promising or frequently visited branches, which is a well-known issue in MCTS design \cite{mcts2012survey}, potentially overlooking branches that contain critical but infrequent behaviors.

Compared to the recent CBC search strategy, \ourSol outperforms it for two reasons.
First and most importantly, the assumption that many paths have no contribution to branch coverage and thus bug finding held by CBC is not always valid for memory error detection, as most of the memory errors can only be triggered under specific contexts (e.g., function calls). The vulnerable function may trigger the bug in one function call, but invoking another function call may not.
Second, computing dependence information on demand in CBC takes time and memory consumption (as reported in the CBC paper \cite{yi2024compatible}, it introduces 30\% more memory consumption and execution time than the baseline approach KLEE when performing path exploration).
\rev{Compared to \cgs, \featmaker, and \empc, \ourSol also detects most of the unique memory errors, indicating its superior performance.
This is mainly because these approaches all aim to improve code coverage (the concrete branches explored by \cgs; the specific features targeted by \featmaker; and the path cover information utilized by \empc) while being agnostic to vulnerable behaviors, making them less effective at detecting unique memory errors.
Although it may be possible to modify them to be vulnerability-oriented, it can be hard to perform a smart path search that considers both forward and backward program information during path exploration, as does \ourSol (see more detailed comparison and discussion in Section \ref{mcts::sec:related-work}).
\ourSol, on the other hand, directly targets exploring unsafe program paths, thus achieving better performance in both metrics.
}

\begin{resultbox}
    {\bf Answer to RQ1:} \rev{Given a one-hour execution period, Vital surpasses current coverage-guided path search approaches by covering up to 90.03\% of unsafe pointers and uncovering up to 57.14\% more unique memory errors.}
\end{resultbox}

\smallskip
\noindent
\textsc{\textbf{Takeaway:}} {\it We analyze the correlation between the number of unsafe pointers and memory errors and present the result in Figure \ref{mcts::fig:correlation} using the data collected in Table \ref{mcts::tab:rq1}. The statistical Pearson's coefficient {\bf 0.924} suggests a strong positive correlation  \cite{cohen2009pearson}, meaning a higher number of unsafe pointers suggests a greater likelihood of vulnerabilities.}

\subsection{RQ2: Vulnerability Detection Capability}  \label{mcts::sec:evaluation:rq2}

{\bf Benchmarks.}  We use four CVEs in GNU {\tt libtasn1} library with different versions (followed Chopper \cite{trabish2018chopped}), namely CVE-2012-1569 (v3.21), CVE-2014-3467 (v3.5), CVE-2015-2806 (v4.3), and CVE-2015-3622 (v4.4).
The {\tt libtasn1} library facilitates the serialization and deserialization of data using Abstract Syntax Notation One (ASN.1).
The selected vulnerabilities predominantly involve memory out-of-bounds accesses. 
Note that each identified vulnerability is associated with detecting a single failure, except for CVE-2014-3467, where this vulnerability manifests across three distinct locations, so the experiment seeks to detect six distinct vulnerabilities.

\smallskip
{\bf Approaches and Evaluation Metrics for Comparison.} We compared \ourSol with the baseline approach (i.e., KLEE \cite{klee}) and \rev{five} state-of-the-art approaches (i.e., \chopper \cite{trabish2018chopped}, CBC \cite{yi2024compatible}, \rev{\cgs \cite{cgs-icse24}, \featmaker \cite{featMaker-fse24}, and \empc \cite{empc-sp25}}) in this RQ. 
To be specific, following existing studies \cite{yi2024compatible,trabish2018chopped}, we run KLEE, \chopper, and CBC under different search strategies, i.e., {\it Random (\tt random)}, {\it DFS (\tt dfs)}, and {\it Coverage (\tt nurs:covnew)}, while keep other with default search strategy.
We use the following two metrics to assess their effectiveness.

\begin{itemize}[leftmargin=1em,nosep]
    \item (1) {\it Execution time} records the time to detect a vulnerability.
    \item (2) {\it Memory consumption} measures memory usage in detecting a vulnerability.
\end{itemize}

\smallskip
{\bf Running Setting.} Following the existing study \cite{trabish2018chopped}, we set a timeout of 24 hours to detect each vulnerability.
Again, for random searches, we ran them five times and reported the median results.

\smallskip
{\bf Results.}   \label{mcts::sec:evaluation:rq2:memory-usage}
Table \ref{mcts::tab:rq2} presents the overall results in terms of execution time. 
From the table, we can see that \ourSol outperforms \chopper and CBC for almost all vulnerabilities. 
Upon detecting the vulnerability in CVE-2014-3467$^2$, \ourSol achieves a speedup of 30x to \chopper.
For the vulnerability CVE-2015-2806, \ourSol takes slightly more time (less than 20 seconds) to detect it.
This is because \chopper skips some large functions before execution, making it detect the vulnerability faster.
However, as emphasized in Section \ref{mcts::sec:introduction}, users of \chopper need prior expert knowledge to decide which functions/lines to skip, which could be a labor-intensive and time-consuming task. 
Also, \chopper consumes more memory when detecting the vulnerability (see more details below).

\begin{table*}[t]
    \footnotesize
    \centering
    \caption{\rev{Results of time on detecting known CVE vulnerabilities.}}
    \vspace{-1em}
    \begin{tabular}{|l|l|c|c|c|c|c|c|}
        \hline
       \multirow{2}{*}{\textbf{Approaches}} & \multirow{2}{*}{\textbf{Search}} & \multirow{2}{*}{\textbf{2012-1569}} & \multirow{2}{*}{\textbf{2014-3467$^1$}} & \multirow{2}{*}{\textbf{2014-3467$^2$}} & \multirow{2}{*}{\textbf{2014-3467$^3$}} & \multirow{2}{*}{\textbf{2015-2806}}  & \multirow{2}{*}{\textbf{2015-3622}}\\
        & & & & & & &  \\
         \hline
        \multirow{3}{*}{KLEE \cite{klee}} & {\tt random} & 00:11:36 & {\it Time-out} & 00:00:06 & {\it Time-out} & 00:05:43 & {\it Time-out} \\
        & {\tt dfs} & 00:02:40 & 00:03:07 & 14:55:21 & {\it Time-out} & 02:55:37 & {\it Time-out} \\
        & {\tt coverage} & 00:11:03 & OOM & 00:00:05  & {\it Time-out} & OOM & 07:49:39\\
        \hline
        \multirow{3}{*}{\chopper \cite{trabish2018chopped}} 
        & {\tt random} & 00:01:50 & 00:08:24 & 00:09:25 & 00:26:48 & 00:02:33 & 00:00:58\\
        & {\tt dfs} & 00:01:03 & 00:00:12 & 00:58:48 & 00:00:29  & {\it Time-out} & 00:13:56 \\
        & {\tt coverage} & 00:01:57 & 00:04:22 & 00:19:11 & 00:11:22 & 00:01:48 & 00:00:57 \\
        \hline
        \multirow{3}{*}{CBC \cite{yi2024compatible}} 
        & {\tt random} & 00:01:01 & {\it N/A} & 00:00:29 & 00:00:54 & {\it N/A} & 00:02:40 \\
        & {\tt dfs} & 00:00:48 & {\it N/A} & {\it Time-out}   & 00:00:18 & 00:21:48 & 00:01:40\\
        & {\tt coverage} & 00:01:18 & {\it N/A} & 00:00:23 & {\it Time-out}  & {\it N/A} & 00:01:19 \\
        \hline 
        \multirow{1}{*}{\rev{\cgs} \cite{cgs-icse24}} & \rev{{\tt cgs}} & \rev{00:11:18} & \rev{18:18:09} & \rev{00:00:14}  & \rev{00:33:40} & \rev{00:50:29} & \rev{00:10:01}\\
        \hline
        \multirow{1}{*}{\rev{\textsc{Feat.}} \cite{featMaker-fse24}} & \rev{{\tt featmaker}} & \rev{00:27:23} & \rev{{\it Time-out}} & \rev{00:00:11}  & \rev{00:49:09} & \rev{00:01:24} & \rev{00:32:30}\\
        \hline
        \multirow{1}{*}{\rev{\empc} \cite{empc-sp25}} & \rev{{\tt empc}} & \rev{-} & \rev{OOM} & \rev{00:00:10} & \rev{{\it Time-out}} & \rev{00:17:37} & \rev{00:00:38} \\
        \hline
        \multirow{1}{*}{\ourSol} & \rev{{\tt mcts}} & \rev{00:01:03} & \rev{00:01:06}  & \rev{00:00:19} & \rev{00:00:09} & \rev{00:02:53} & \rev{00:00:26} \\
        \hline
    \end{tabular}
    \label{mcts::tab:rq2}
    \begin{tablenotes}
        \footnotesize
        \item * The time format is {\it hour:minute:second}. ``{\it N/A}'' represents the normal termination of execution without reproducing the vulnerability; ``{\it OOM}'' means Out-of-Memory. {\it Time-out} means the execution exceeded the time limit without reproducing the vulnerability. The index number $n$ in CVE-2014-3467$^{n}$ represents a distinct location of the vulnerability manifested. \rev{The `-' in the row of \empc indicates that \empc cannot successfully reproduce the CVE.}
    \end{tablenotes}
\end{table*}

Compared to CBC\footnote{There are some differences between the results in Table \ref{mcts::tab:rq2} and those in \cite{yi2024compatible} Table 2, mainly due to different settings used.
We use a \emph{fixed} setting that produces the best overall results, instead of the \emph{best setting for each} CVE, which requires expert knowledge about each CVE, and we think would not be a \cc{practical usage}
(details in Appendix-A in Supplementary Material
).}, \ourSol takes less time to detect 5 out of 6 vulnerabilities (except CVE-2012-1569 with {\it Random} and {\it DFS} search). 
For CVE-2014-3467$^1$, CBC does not detect it in any search strategies, as this vulnerability happens in a {while-loop} condition, which requires multiple executions of the condition to trigger the vulnerability. 
This again emphasizes a fundamental deficiency in CBC, as it can only detect bugs about assertion failure that are triggered when the code executes once. 
However, as shown in previous studies \cite{haller2013dowsing,aeg,mayhem}, a plethora of memory errors occur in loop iterations, and only executing the loop once is insufficient to detect the important category of memory errors. 
\rev{Compared to the remaining \cgs, \featmaker, and \empc, \ourSol is still superior, as \ourSol detects all the vulnerabilities while others miss some of them.
For those cases that others can detect, \ourSol also takes less time to detect them in 4 out of 6 cases.}

\begin{figure}[t]
\vspace{-1em}
\centering
\subfigure[CVE-2012-1569]{	
\begin{minipage}[t]{0.316\linewidth}
\centering
\includegraphics[width=1\linewidth]{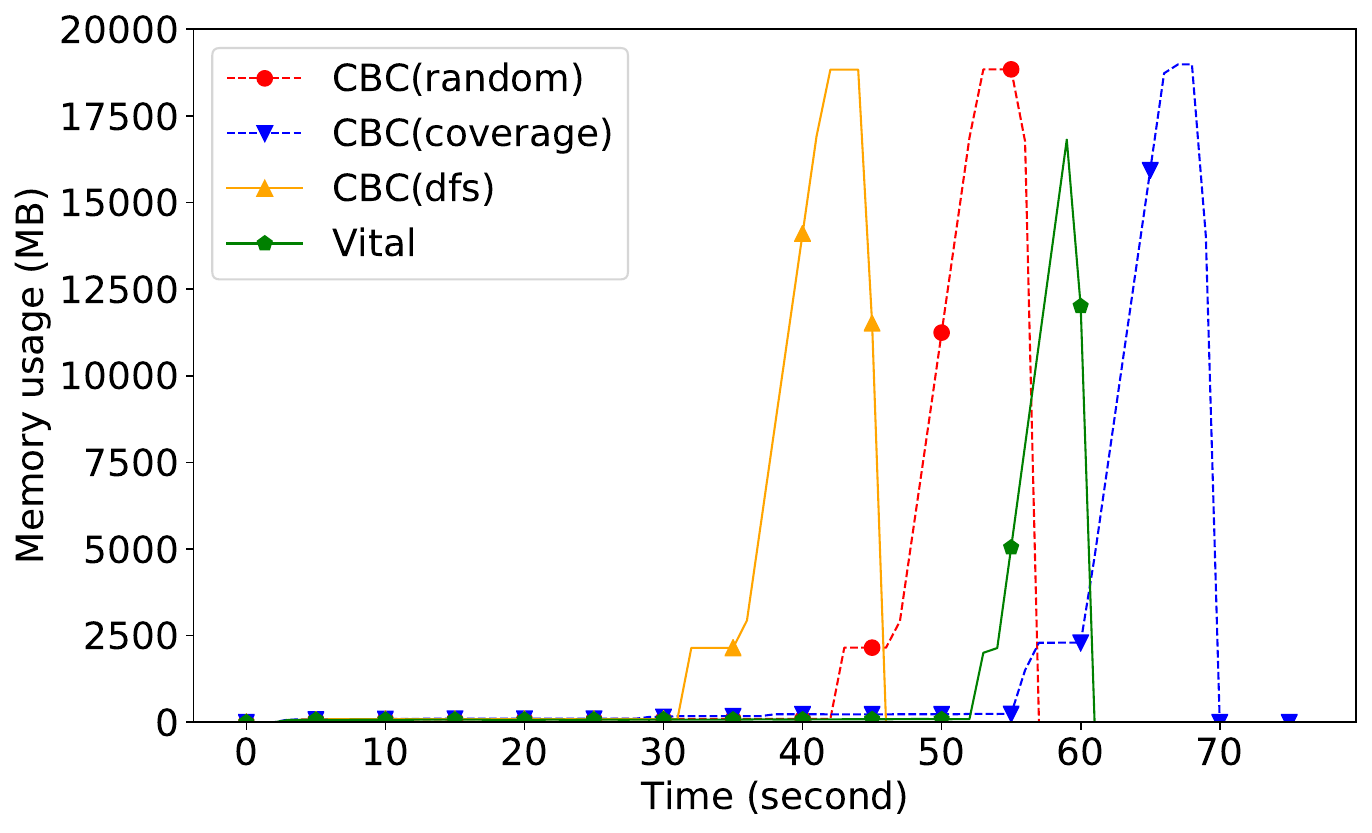}
\vspace{-1em}
\label{mcts::fig:memory-usage-cbc}
\end{minipage}	
}
\subfigure[\rev{CVE-2014-3457$^2$}]{	
\begin{minipage}[t]{0.316\linewidth}
\centering
\includegraphics[width=1\linewidth]{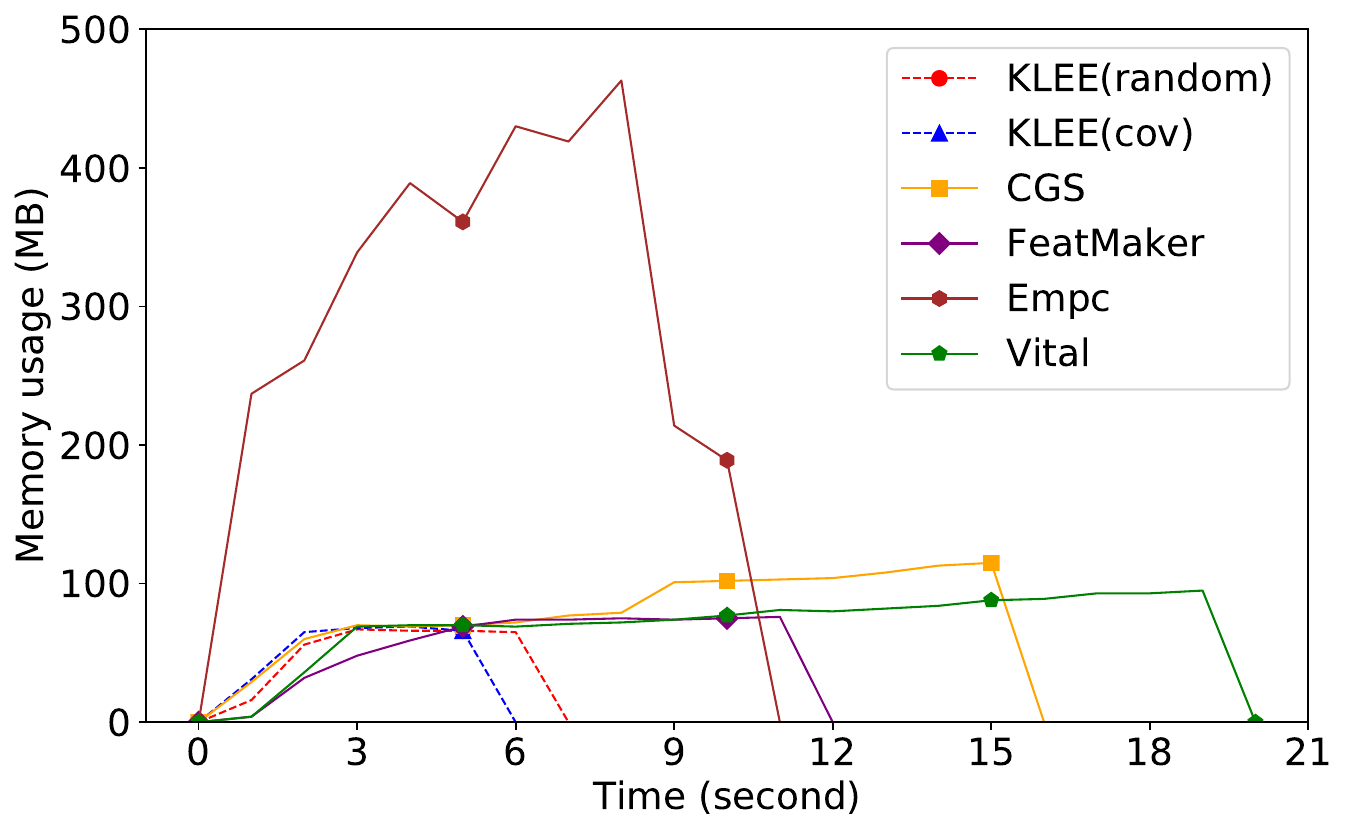}
\vspace{-1em}
\label{mcts::fig:memory-usage-cbc}
\end{minipage}	
}
\subfigure[CVE-2015-2806]{
\begin{minipage}[t]{0.316\linewidth}
\centering
\includegraphics[width=1\linewidth]{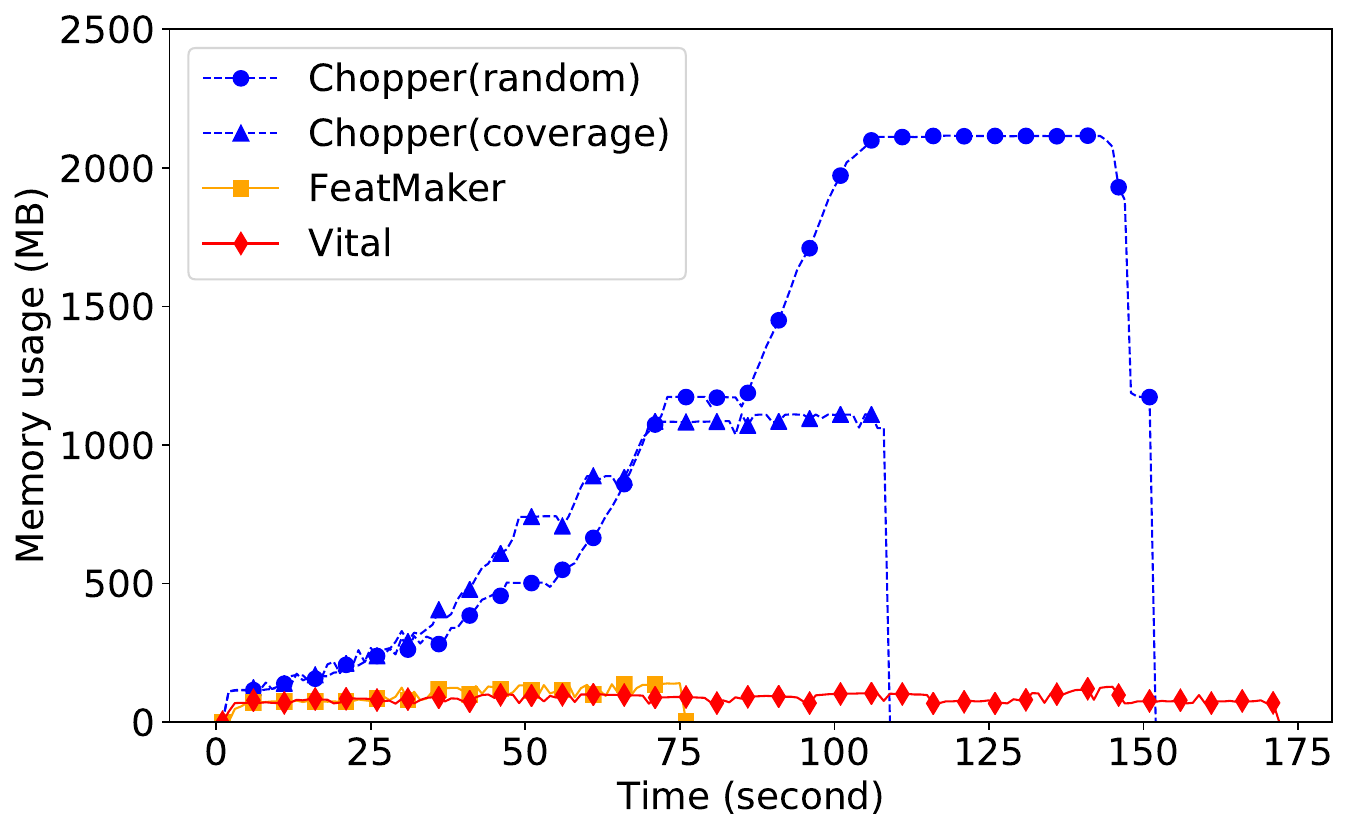}
\vspace{-1em}
\label{mcts::fig:memory-usage-chopper}
\end{minipage}
}
\vspace{-1em}
\caption{\rev{Results of memory consumption (only \ourSol takes comparable or slightly more time are presented)}
}
\label{mcts::fig:memory-usage}
\end{figure}

In terms of memory consumption, we run \ourSol against \chopper and CBC over two vulnerabilities in CVE-2015-2806 and CVE-2012-1569, respectively, to gain more insights because \ourSol takes comparable or slightly more time for the vulnerability detection.
Figure \ref{mcts::fig:memory-usage} shows the results, where {\it x-axis} represents the time to detect the vulnerability and {\it y-axis} indicates the usage of memory.
From the two figures, we can observe that \ourSol consistently consumes less memory when detecting the vulnerability.
In particular, as shown in Figure \ref{mcts::fig:memory-usage}(a), we can observe that \ourSol consumes significantly less memory when detecting the same vulnerabilities compared to \chopper.
For example, \chopper with a random search consumes at most 2,115 MB of memory, while \ourSol only takes around 100 MB of memory, producing a significant reduction (i.e., 20x) in memory consumption. 
This is reasonable as \chopper takes a state recovery mechanism to maintain the execution of skipped functions. Since the skipped functions can be very large, maintaining recovered states in \chopper tends to be memory-intensive.
Compared to CBC, it requires extra memory consumption to obtain the dependence information during the symbolic execution, which is memory-intensive.
In contrast, \ourSol does not increase memory consumption compared to standard symbolic execution, providing a solution that is not only lightweight but also highly effective.
\rev{Compared to \empc over CVE-2014-3467$^2$, although \ourSol takes slightly more time than \empc, \ourSol consumes 5x less memory usage, demonstrating its efficiency.}

\rev{The superior performance from Vital is expected, as existing approaches all aim to improve code coverage while being agnostic to vulnerable behaviors, making them less effective at detecting vulnerabilities.
Specifically, \cgs focuses on exploring concrete branches, which may not necessarily lead to the exploration of paths that trigger vulnerabilities.
\featmaker targets specific features, which may not align with the paths that expose vulnerabilities.
\empc utilizes path cover information, which may not prioritize paths that lead to vulnerabilities.
In contrast, \ourSol directly targets exploring unsafe program paths, thus achieving better performance.
}

\begin{resultbox}
 {\bf Answer to RQ2:} \ourSol outperforms chopped symbolic execution by achieving a speedup of up to 30x execution time and a reduction of up to 20x memory consumption.
\end{resultbox}

\subsection{RQ3: Ablation Studies}  \label{mcts::sec:evaluation:rq3}

\smallskip
{\bf Benchmarks and Evaluation Metrics.}  We use the same benchmarks and evaluation metrics as in RQ1 to compare different variant approaches of \ourSol to answer RQ3 in this subsection.

\smallskip
{\bf Variant Approaches.} We design several variants of \ourSol to gain a deeper understanding of the contribution of each component. We focus on evaluating the following variant approaches:
\begin{itemize}[leftmargin=1em,nosep]
    \item \ourSolNoExp uses random expansion without guided expansion.
    \item \ourSolNoSim performs path search without simulation. 
    \item \ourSolNoSimOpt simulates without \cc{the optimization for loops}.
    \item \textsc{Vital(OPT2-x)}, where x represent the number of instructions to terminate the simulation of {\it OP2} (as discussed in Section \ref{mcts::approach:simulation}, {\it OP2} terminates the simulation when x instructions are executed).
\end{itemize} 

\smallskip
{\bf Running Setting:} We use the same setting as in RQ1: each approach with a one-hour run and use the metrics (i.e., the number of unsafe pointer coverage and memory errors detected) to evaluate their effectiveness.

\smallskip
{\bf Results.} 
Table \ref{mcts::tab:rq3} presents the overall results.
We can conclude that \ourSol performs better than all variant approaches.
Specifically, compared to \ourSolNoExp, \ourSol could cover 16.80\% more unsafe points and detect 7.32\% more memory errors, demonstrating the contribution of unsafe pointer-guided node expansion designed in Algorithm \ref{mcts::alg:mcts-overall}.
The superior performance of \ourSol is reasonable as random expansion tends to waste exploration efforts on non-vulnerable paths, thus downgrading the overall performance of MCTS.
The other two variant approaches share similar results, where \ourSol achieves an improvement of 43.04\% and 22.22\% as well as 30.67\% and 33.33\% in terms of the number of covered unsafe pointers and detected memory errors, compared to \ourSolNoSim and \ourSolNoSimOpt, respectively, indicating the contribution of simulation and its optimization.
The results of \ourSolNoSim justify that without simulation, MCTS cannot fully utilize the past execution information to guide the path exploration, thus downgrading its performance.
The results of \ourSolNoSimOpt also demonstrate the effectiveness of the loop optimization in improving the simulation performance, as loops are widely used in real-world programs.

\smallskip
{\bf Evaluation of Different Simulation Strategies}.
We run {\it OP3} (the default simulation setting in \ourSol), \ourSolOptTwoF, \ourSolOptTwoO, and \ourSolOptTwoT (where x in \textsc{Vital(OPT2-x)} means the fixed number \cc{of instructions} to terminate the simulation) to evaluate different settings of {\it OP2}. 
We omit {\it OP1} as its static simulation of a node tends to be ineffective due to time-consuming CFG traversing and high false alarm rate, as discussed in Section \ref{mcts::approach:simulation}.
As shown in Table \ref{mcts::tab:rq3}, all three settings of {\it OP2} perform inferior compared to {\it OP3} (normally terminated execution) in \ourSol.
This is reasonable as aborting too early would make \ourSol prone to false negatives, thus missing interesting unsafe pointers coverage and downgrading the overall performance of MCTS.

\smallskip
\rev{{\bf Comparison with Unsafe Pointers Guided Search without MCTS.}}
\rev{Another simpler variant approach is to use the indicator (i.e., type-unsafe pointers) to guide the search without using MCTS. Theoretically, the simpler heuristic is limited for two reasons. First, like many existing search techniques (as thoroughly discussed in Table \ref{related-work::tab:approach-comparison} and in Section \ref{mcts::sec:related-work}), the simpler heuristic is agnostic to the past execution, which is shown to be an important factor to boost the path exploration. 
Second, the limited capability of balancing the exploration of future unexplored paths and the exploitation of the past explored paths during path search makes it costly and difficult to reach the known location of type-unsafe pointers. 
In contrast, Vital is superior for both aspects: MCTS can utilize the past execution information to guide the path exploration, and it has a good potential to balance the exploration and exploitation during path search, making Vital an optimal solution towards the vulnerability-guided path exploration.
}

\begin{resultbox}
{\bf Answer to RQ3:} The newly designed components, including type-unsafe pointer-guided node expansion and node simulation as well as their loop optimization, are imperative to boost the performance of \ourSol in terms of covering unsafe pointers and detecting memory errors.
\end{resultbox}

\begin{table}[t]
    \centering
    \small
    \caption{Comparison results with variant approaches}
    \vspace{-1em}
    \begin{tabular}{lrr|rr}
        \hline
        \multirow{2}{*}{\textbf{Search}} & \multicolumn{2}{c|}{{\bf Unsafe Pointers}} & \multicolumn{2}{c}{{\bf Memory Errors}}  
         \\
        \cline{2-5}
        & \textit{Num$_{total}$} & \textit{Imp$_{total}$} & \textit{Num$_{total}$} & \textit{Imp$_{total}$} \\
        \hline
        \ourSolNoExp & 11034 & 16.80\%  & 41 & 7.32\%   \\
        \ourSolNoSim & 9010  & 43.04\% & 36& 22.22\%   \\
        \ourSolNoSimOpt & 9863 & 30.67\% & 33 & 33.33\%   \\
        \hline
        \ourSolOptTwoF & 5449 & 137.52\% & 25 & 76.00\%  \\
        \ourSolOptTwoO & 7805 & 65.12\% & 26 & 69.23\% \\
        \ourSolOptTwoT & 6477 & 98.98\% & 26 & 69.23\% \\
        \hline
        \ourSol & 12888  & - & 44 & -  \\
        \hline
    \end{tabular}
	\label{mcts::tab:rq3}
\end{table}

\subsection{RQ4: Practical Application of \ourSol}  \label{mcts::sec:evaluation:rq4}

To demonstrate the practical vulnerability detection capability of \ourSol over large-scale software systems, we run \ourSol over the intensively-tested and latest released {\it objdump} package (includes more than 20k lines of code) in {\tt GNU binutils-2.44}.
As a result, \ourSol detected a previously unknown vulnerability\footnote{\url{https://sourceware.org/bugzilla/show_bug.cgi?id=32716}.} within one minute. 
The issue is a memory leak where an allocated object is not appropriately de-allocated, which may cause severe security risks such as critical information leakage or denial of service.
This issue had been lurking for more than 7 years before we reported it\footnote{The oldest version that can reproduce the issue is binutils-2.29 released on 25/09/2017.}. 
Since the critical impact of the issue, developers confirmed and fixed it swiftly within 8 hours after we submitted the bug report. 
\cc{It has been assigned a new CVE ID (i.e., CVE-2025-3198).}

It is worth noting that other approaches face difficulties in detecting this vulnerability.
To have a better understanding, we ran KLEE, \chopper, and CBC on the same version of {\it objdump} for 24 hours but it turns out that none of them could detect this issue\footnote{We provide detailed vulnerability analysis in Appendix-B in the Supplementary Material.} 
In summary, KLEE failed to cover the vulnerable path because of the complex input conditions and multiple loop iterations required to reach the vulnerable function.
\chopper \cite{trabish2018chopped} failed to detect the issue mainly due to the lack of prior expert knowledge or technical support to set up the execution. KLEE's default search heuristics are limited by bypassing the input-dependent loops, which are prerequisites to reach the target vulnerable function.
CBC \cite{yi2024compatible} is restricted by the non-vulnerability-oriented path search heuristic, thus missing the vulnerability: it ignores the fact that many memory errors can only happen after a fixed number of executions on certain loops \cite{haller2013dowsing}.

\ourSol covers the vulnerable path because of its new way of simulating execution paths
and its novel search based on MCTS.
Compared with existing solutions, \ourSol can skip {\it unimportant} loops that have no contribution to the accumulation of unsafe pointer coverage by using simulation optimization strategies (see more details in Section \ref{mcts::approach:simulation}).
For loops that may lead to memory issues, our simulation and MCTS will evaluate the reward of each of their iterations and continue when new unsafe pointers are covered.

\begin{resultbox}
    {\bf Answer to RQ4:} \ourSol is able to detect previously unknown vulnerabilities in practice, working in a fully automatic manner without the need for prior expert knowledge.
\end{resultbox}

\section{Discussion}  \label{mcts::sec:discussion}


\smallskip
\textbf{Overhead of Pointer Type Inference.} \label{mcts::sec:discussion:ccured-time}
To further understand the overhead of the type inference system (i.e., \ccured) used in \ourSol, we measure the time to infer pointer types.
The results show that in most cases (64\%, 48 out of 75), the time spent on the analysis is within 5 seconds. 
We believe that such overheads are negligible compared to the large amount of time used for the entire testing period (e.g., 24 hours in RQ2).


\smallskip
\textbf{Impact of Different Configurations.} \label{mcts::sec:discussion:impact}
The selection of the value of the parameter {\tt C} defined in Equation in Section \ref{mcts::approach:selection} and  ``{\it --optimization-degree}'' in Algorithm \ref{mcts::alg:doSimulation} may affect the performance of \ourSol.
Thus, we conduct extra experiments to assess the impact of different running configurations.

For the 
parameter {\tt C} in Section \ref{mcts::approach:selection}, we run \ourSol with values of \ $\sqrt[]{2}$, 5, 10, 20, 50, and 100. 
The results show that \ourSol covered 12,888, 12,620, 13,144, 12,724, 13,188, and 12,578 unsafe pointers and detected 44, 41, 42, 43, 43, and 44 memory errors in each configuration, respectively. Since the goal of \ourSol is to detect more memory vulnerabilities, we select the value \ $\sqrt[]{2}$ as the default configuration of \ourSol as this setting yields the best memory error detection and comparable unsafe pointer coverage capabilities.

For the impact of ``{\it --optimization-degree}'', we run \ourSol with values of 100, 300, 500, 700, 1{,}100, and 1{,}500.
The results show \ourSol covered 12{,}078, 12,349, 12,660, 12,888, 12,655, and 12,578 unsafe pointers and detected 35, 39, 42, 44, 41, and 39 memory errors under each configuration, respectively. Since the value of 700 produces the best results in terms of both unsafe pointer coverage and memory error detection capabilities, we set 700 as the default value in \ourSol.

\smallskip
\textbf{Threats to Validity.} \label{mcts::sec:discussion:threats}
The {\it internal} validity concerns stem from the implementation of \ourSol. To mitigate this threat, we built \ourSol on top of the well-maintained and recently released version (v3.0) of KLEE \cite{klee}. In addition, we have meticulously implemented \ourSol, reusing existing APIs in KLEE, as explained in Section \ref{mcts::sec:implementation}, and have performed a detailed code check to mitigate the threat. 

The {\it external} threat comes from the benchmarks used in this study. We used GNU \texttt{Coreutils} and a library {\tt libtasn1} with four different versions. Although they have been widely used for evaluating symbolic execution \cite{klee,ccs21-learch,steering-se,trabish2021bounded,trabish2020relocatable}, these programs may not be representative enough for various software systems. To further alleviate these potential threats, we are committed to expanding the program sets in our future work.
To mitigate such a threat, we also evaluated \ourSol on a larger test program in RQ4 and detected an unknown vulnerability.

The {\it construct} validity threat is subject to configurations of parameters. We address this concern by thoroughly investigating their impact, which enhances transparency and enables a deeper understanding of the influence of different configurations. 
Following many existing studies \cite{nagarakatte2009softbound,hardbound,cguard-issta23,sticktags-sp24,tailcheck}, another threat of \ourSol is that the current version supports ensuring spatial memory safety only.  
This is mainly because the tool used for type inference \ccured \cite{ccured-popl2002} can not classify all unsafe pointers (i.e., the ones that lead to {\it temporal} memory errors). 
Note that extending the support to more types of {\it new} indicators to detect more types of vulnerabilities in \ourSol should only involve engineering effort (e.g., transporting the tool SAFECode \cite{safecode} to work on compatible LLVM bitcode to collect {\it unsafe} pointers to help detect temporal memory errors). 
We plan to alleviate this threat in future work.

\smallskip
\textbf{Limitations.} \label{mcts::sec:discussion:limitation}
\ourSol suffers from certain inherent limitations in symbolic execution engines (e.g., KLEE \cite{klee}) due to limited memory modeling, environment modeling, and efficiency issues, which may restrict the memory error detection capability of \ourSol. This is because some intractable vulnerabilities can only be triggered under a complex situation, which requires a more comprehensive modeling of the program semantics of test programs. 
Efficiency is also an issue, as most symbolic execution engines analyze test programs by interpreting the intermediate representation code (e.g., LLVM Bitcode in KLEE), which is shown to be inefficient \cite{symcc,pitigalaarachchi2023krover}. Recent studies \cite{symloc,tu2022fastklee,schemmel2023kdalloc,pandey2019deferred} have been proposed to resolve these problems, and we plan to integrate them in future work.

\rev{Following many studies \cite{nagarakatte2009softbound,hardbound,cguard-issta23,sticktags-sp24,tailcheck}, another limitation of \ourSol, which we inherit from the static analysis tool \ccured \cite{ccured-popl2002}, is that the current \ourSol ensures {\it spatial} memory safety only. 
This is mainly because \ccured, the tool used for type inference, can not classify all unsafe pointers (e.g., the ones leading to {\it temporal} memory errors).
There are several directions to address this limitation.
First, one may add the indicators of temporal memory errors (e.g., dangling pointers) to guide the path exploration in \ourSol.
Note that extending the support to more types of {\it new} indicators to detect more types of vulnerabilities in \ourSol should only involve engineering effort (e.g., transporting the SAFECode tool \cite{safecode} to work on compatible LLVM bitcode to collect {\it unsafe} points to help detect temporal memory errors). 
We plan to leverage more advanced techniques (such as those proposed in DataGuard \cite{dataGuard}) to address this limitation in future work.
}
\rev{Second, one may employ advanced static analysis tools such as Infer \cite{infer}, which offers comprehensive detection of memory safety issues (including temporal memory), to guide path exploration.
However, unlike \ccured, which has the guarantee of soundness (i.e., free of false negatives) by design, Infer and other tools may suffer from both false positives and negatives \cite{cui2024empirical,lipp2022empirical}.
To enable a more robust cooperation with other static analysis tools beyond \ccured, one has to design an appropriate solution to reduce false positive and negative rates before using their results to guide the path exploration for symbolic execution. We consider such a direction for future work as well.
}

\section{Related Work} \label{mcts::sec:related-work}

{\bf Techniques for Alleviating Path Explosion.}
Various techniques are introduced to tackle the path explosion problem; the most relevant include path search strategies and under-constrained (e.g., {\it chopped}) symbolic execution.
Most search heuristics for symbolic execution are coverage-guided.
Cadar {\it et al.} \cite{cadar2008exe} propose a Best-First Search strategy. 
Cadar {\it et al.} further propose KLEE \cite{klee}, where random and code coverage-guided ({\tt nurs:covnew}) search strategies are proposed. Later, KLEE continued to upgrade to support many more strategies, such as {\tt bfs}, {\tt dfs}, and instruction coverage-guided ({\tt nurs:md2u} and {\tt nusrs:icnt}).
Burnim {\it et al.} \cite{burnim2008heuristics} propose using a weighted control flow graph (CFG) to guide exploration to the nearest uncovered parts based on the distance in the CFG.
Li {\it et al.} \cite{steering-se} propose to exploit a new {\it length-n} subpath program spectra to systematically approximate full path information for guiding path exploration.
He {\it et al.} \cite{ccs21-learch} adopt a machine learning-based strategy to first train a model based on program execution information and then effectively select promising states with the trained model.
\rev{\textsc{UbSym} \cite{ubsym} exploits test units that are relevant to vulnerability to prioritize the path exploration in binary programs.}
CBC \cite{yi2024compatible} proposes a compatibility-based search strategy to prioritize the execution states that are more likely to be compatible with other states, thus increasing the chance of generating new paths.
\rev{
\cgs \cite{cgs-icse24} proposes a concrete-path search strategy to prioritize the execution states that can lead to new paths through concrete execution.
\featmaker \cite{featMaker-fse24} proposes a feature-based (i.e., path conditions) search strategy to prioritize the execution states that can maximize the performance of symbolic execution in terms of code coverage and bug detection.
\empc \cite{empc-sp25} proposes a new path exploration strategy based on minimum path cover (MPC), where MPC provides an option for runtime path selection to use the least number of paths to maximize code coverage.
}
Unlike conducting a path search only for code coverage, there are only a few vulnerability-oriented search strategies.
{\tt StatSym} \cite{yao2017statsym} instruments test programs to construct predicates that indicate vulnerable features and then employs a path construction algorithm to select the vulnerable paths.
{\tt SyML} \cite{ruaro2021syml} guides path exploration toward vulnerable states through pattern learning. 
However, existing coverage-guided techniques give the same priority to code that is unlikely to contain vulnerabilities. In contrast, \ourSol maximizes the number of unsafe pointers to increase the likelihood of exposing memory unsafe vulnerabilities.
Furthermore, vulnerability-oriented approaches require a training set of vulnerabilities previously discovered in the program and try to link patterns in the runtime information to vulnerabilities. In contrast, \ourSol requires neither training nor unspecific runtime information to quantify the vulnerability-proneness of a path.
Since both {\tt StatSym} and {\tt SyML} are not open-sourced and {\tt SyML} only focuses on binary programs, we could not directly experimentally compare them in this paper.

Regarding  under-constrained symbolic execution,
Csallner {\it et al.} \cite{csallner2005check} and Engler {\it et al.} \cite{engler2007under} propose and extend the idea of under-constrained symbolic execution, where the symbolic executor cuts the code (e.g., an interesting function) to be analyzed out of its enclosing system and checks it in an isolation manner. 
Recent work \chopper \cite{trabish2018chopped} cuts out uninteresting functions that are vulnerability irrelevant.

\rev{To help readers better understand the differences between \ourSol and existing path exploration techniques, we summarize the main differences between the techniques compared in the evaluation and \ourSol in Table \ref{related-work::tab:approach-comparison}.
We compare them from four perspectives: (1) whether the technique is vulnerability-oriented; (2) whether the technique is fully automated without requiring expert knowledge; (3) whether the technique can learn from past execution information to improve path exploration, and (4) whether the technique can balance exploration of {\it unexplored} paths and exploitation of {\it explored} paths during path search.
As shown in Table \ref{related-work::tab:approach-comparison}, \ourSol is the only technique that supports all four features to improve path exploration.
Specifically, compared with existing representative coverage-guided or other heuristics for path exploration (i.e., KLEE \cite{klee}, CBC \cite{yi2024compatible}, \cgs \cite{cgs-icse24}, \featmaker \cite{featMaker-fse24}, and \empc \cite{empc-sp25}), \ourSol is the only one that is vulnerability-oriented, can learn from past execution information, and can balance exploration and exploitation to improve path exploration.
Furthermore, compared with under-constrained symbolic execution, \chopper requires users to specify which functions to skip based on their expert knowledge and cannot learn from past executions.
In contrast, \ourSol is fully automated without requiring expert knowledge.
In summary, compared to existing path search strategies, \ourSol is a novel technique that can automatically perform vulnerability-oriented path exploration using a novel MCTS-based approach that not only learns from past execution but also balances exploration and exploitation during path search, making it advanced in vulnerability-oriented path exploration in symbolic execution.
}

\begin{table}[t]
	\centering
    \small
	\caption{\rev{Comparison of different path exploration methods (\ding{52} represents holds the feature while \ding{56} does not).}}
    \vspace{-1em}
	\begin{tabular}{cccccccc}
		\toprule
        \multirow{1}*{\textbf{Features}} & 
		\multirow{1}*{KLEE \cite{klee}} & 
        \multicolumn{1}{c}{Chopper \cite{trabish2018chopped}} &
	\multicolumn{1}{c}{CBC \cite{yi2024compatible}} &
        \multicolumn{1}{c}{\cgs \cite{cgs-icse24}} &
        \multicolumn{1}{c}{\featmaker \cite{featMaker-fse24}} &
        \multicolumn{1}{c}{\empc \cite{empc-sp25}} &
	\multicolumn{1}{c}{\ourSol} \\
		\midrule
        \multirow{1}{*}{{\bf F1}} & \ding{56}  & \ding{52} & \ding{56} & \ding{56} & \ding{56} & \ding{56} & \ding{52} \\
		\multirow{1}{*}{{\bf F2}} & \ding{52}  & \ding{56} & \ding{52} & \ding{52} & \ding{52} & \ding{52} & \ding{52}  \\
		\multirow{1}{*}{{\bf F3}} & \ding{56}  & \ding{56} & \ding{56} & \ding{56} & \ding{56} & \ding{56} & \ding{52} \\
		\multirow{1}{*}{{\bf F4}} & \ding{56}  & \ding{56} & \ding{56} & \ding{56} & \ding{56} & \ding{56} & \ding{52} \\
		\bottomrule
	\end{tabular} 
    \begin{tablenotes}
        \footnotesize
        \item * {\bf F1}: vulnerability-oriented? {\bf F2}: fully automated? {\bf F3}: learn from past execution? {\bf F4}: balance exploration \& exploitation?
    \end{tablenotes}
	\label{related-work::tab:approach-comparison}
\end{table}

\smallskip
{\bf Applications of Monte Carlo Tree Search.}
MCTS was initially applied to enhance heuristics for a theorem prover \cite{ertel1989learning}. Then, it was used to optimize program synthesis \cite{lim2016field} and symbolic regression \cite{white2015programming}. 
Furthermore, MCTS is employed in Java PathFinder \cite{poulding2015heuristic}, where it was used for explicit state model checking.
Liu {\it et al.} \cite{liu2020legion} adopt MCTS to achieve the best trade-off between concolic execution and fuzzing for coverage-based testing. 
Zhao {\it et al.} \cite{zhao2022alphuzz} model the seed scheduling problem in fuzzing as a decision-making problem and use MCTS to select optimal seeds.
The works most related to ours are the {\tt canopy}  \cite{luckow2018monte} and the approach proposed by Yeh {\it et al.} \cite{yeh2017path}.
{\tt canopy} uses MCTS to guide the search for costly paths, where the cost is defined as memory consumed and execution time along a path. 
Yeh {\it et al.}' approach utilizes MCTS to select valuable paths, where the valuable refers to the number of visited basic blocks.

\ourSol adopts MCTS for vulnerability-oriented path exploration for symbolic execution. Compared with the most related works, the differences are as follows. 
(1) Our goal is to detect vulnerable paths in C/C++ programs, while {\tt canopy} aims to find the costly paths in Java programs, and Yeh {\it et al.}' approach focuses on the path with the largest executed number of basic blocks in binary programs.
(2) The node expansion designed in \ourSol is guided by the results of static program analysis, that is, type inference, while both the two compared approaches apply a random expansion strategy.
(3) The simulation policy designed by \ourSol is optimized by previous simulation outcomes, while {\tt canopy} adopts random simulation with limited optimizations and Yeh {\it et al.}' approach uses CFG of the binary program to perform simulation (which may yield imprecise reward, as recovering CFG from binary programs is an undecidable problem \cite{angr}). In summary, \ourSol tends to be more applicable by design to detect new vulnerabilities in practice.

\section{Conclusion with Future Work}  \label{mcts::sec:conclusion}

We present \ourSol, a new vulnerability-oriented symbolic execution via type-unsafe pointer-guided Monte Carlo Tree Search.
\ourSol guides the path search toward vulnerabilities by (1) acquiring {\it type-unsafe} pointers by a static pointer analysis (i.e., type inference), and (2) navigating an optimal exploration-exploitation trade-off to prioritize program paths where the number of unsafe pointers is maximized, leveraging unsafe pointer-guided Monte Carlo Tree Search.
We have compared \ourSol with existing path search strategies and chopped symbolic execution, and the results demonstrate the superior performance of \ourSol among existing approaches in terms of unsafe pointer coverage and memory errors/vulnerabilities detection capability, specifically for practical vulnerability detection capability.
For future work, we are actively pursuing to strengthen vulnerable behavior analysis and modeling in \ourSol.

\section{Data Availability}

The replication package, including the source code and setup instructions for \ourSol, benchmarks, and scripts to reproduce the experiments, is available at {\it \textcolor{mymauve}{\url{https://github.com/haoxintu/Vital-SE}}}.

\begin{acks}
We appreciate Cristian Cadar, Martin Nowack, and Daniel Schemmel for their constructive insights in the earlier stages of this project.
We also thank anonymous reviewers who provided their insightful comments on the previous versions of the draft and the developers who promptly confirmed and fixed the bug we reported. 
\end{acks}

\bibliographystyle{ACM-Reference-Format}
\bibliography{reference}

\appendix

\newpage
\section*{Appendix: Supplementary Material}

\section{Detailed Evaluation Results of CBC \cite{yi2024compatible} on CVE Vulnerability Reproduction}   \label{appendix:B}
Since the experimental results reported by the original paper (Table 2 in CBC \cite{yi2024compatible}) use {\it fine-tuned} settings (the values for their customized ``--reverse-limit'' and ``--states-limit'' options) for every vulnerability in every search strategy, we think such a running setting might be an unfair comparison. To reduce fairness threats, we used the same various values used in the CBC paper for the two options and conducted thorough experiments on those values to investigate their impacts. We set a timeout of 1 hour and record the time for every vulnerability per search strategy.

Table \ref{table:cbc} presents the detailed results. We can observe that there are large variations in the results under different settings. We selected the settings (``--reverse-limit=5'' and ``--states-limit=80'') that can produce the best overall results and reported them in this paper. For the {\it Timeout} cases, we performed another run with an increased timeout of 24 hours for a fair comparison.

\section{Detailed Analysis of the New Vulnerability}   \label{appendix:C}

\subsection{Vulnerability Details}
The simplified vulnerable code snippets adopted from {\tt objdump.c} are shown in Listing 1, where a memory object {\it param->info} (Line 60) is leaked due to improper memory management (i.e., it is allocated but never de-allocated when the execution terminates)\footnote{Full bug report: \url{https://sourceware.org/bugzilla/show_bug.cgi?id=32716}}. The vulnerable point happens in the function {\tt do\_display\_target} starting from Line 52, which is initially invoked by the caller function {\tt display\_info} (Line 34) in the {\tt main} function. After a deeper investigation of the execution flow of the memory leak, we summarized three requirements that a symbolic execution engine should satisfy to make the execution reach the vulnerable point.

\begin{lstlisting}[escapechar=@,caption={Vulnerable code snippets (memory leak at Line 60) from {\tt objdump}},captionpos=b][t] @\label{mcts::listing}@
static bool formats_info;
int main (int argc, char **argv) {
  int c; bool seenflag = false;

  expandargv (&argc, &argv);

  while ((c = getopt_long (argc, argv,
			   "CDE:FGHI:LM:P:RSTU:VW::Zab:defghij:lm:prstvwxz",
			   long_options, (int *) 0)) != EOF) {
    switch (c) {
      case 0: break;		
      case 'm': mainchine = optarg; break;
      case 'M':{
        char *options; //usafe pointer
        if (disassembler_options)
            options = concat (disassembler_options, ...); 
        else
            options = xstrdup (optarg);
        free (disassembler_options);
        disassembler_options =
                remove_whitespace_and_extra_commas (options);
        if (!disassembler_options)
            free (options); //usafe pointer
	  }
	  break; }
      // ...
      case 'i':
        formats_info = true;
        seenflag = true;
        break;
    }
  }
  if (formats_info)
    exit_status = display_info ();
  else {
      ...//
    }
} /* adapted from binutils/objdump.c */

int display_info (void) {
  struct display_target arg;
  display_target_list (&arg);
  ...//
  return arg.error;
} /* adapted from binutils/objcomm.c */

static void display_target_list (struct display_target *arg) {
  .../ 
  bfd_iterate_over_targets (do_display_target, arg);
} /* adapted from binutils/objcomm.c */

static int do_display_target (const bfd_target *targ, void *data){
  struct display_target *param = (struct display_target *) data;
  param->count += 1;
  size_t amt = param->count * sizeof (*param->info);
  if (param->alloc < amt){
    size_t size = ((param->count < 64 ? 64 : param->count)
		* sizeof (*param->info) * 2);
        
    param->info = xrealloc (param->info, size); // @\red{{\bf leaked object}}@
    ...//
  }
  ... //
} /* adapted from binutils/objcomm.c */
\end{lstlisting}

\begin{itemize}[leftmargin=1em,nosep]
    \item \#1. The engine should successfully bypass the function {\tt expandargv} (includes input-independent loops) at Line 5.
    \item \#2. The engine should successfully return from the function {\tt getopt\_long} (includes input-independent loops and requires multiple executions over a loop) in the while-loop (Line 7).
    \item \#3. The return value from the function {\tt getopt\_long} should be the character `i', so that the value of {\it formats\_info} (defined as global variable at Line 1, assigned in a {\it switch-case} statement at Line 28, and used in the {\it if} condition at Line 33) should be {\it true}. 
\end{itemize} 

It would be worth noting that only if the above requirements are satisfied, the branch of {\it if}-condition could be {\it true} at Line 33, and the vulnerable function {\tt display\_info} could be invoked to trigger the memory leak issue. 

\subsection{Why Comparative Approaches Missed it?}
Existing path search strategies failed to detect the issue mainly due to the lack of prior expert knowledge or technical support to set up chopped symbolic execution, limited handling of input-dependent loops (not shown due to page limit), and a restricted (i.e., not vulnerability-oriented) guided path search heuristic. Due to the above reasons, the comparative tools failed to detect the issue by giving a running timeout of 24 hours. 

To set up \chopper \cite{trabish2018chopped}, users need to specify the skipped functions to detect possible vulnerabilities, which could be extremely hard, as users have no idea how to select skipped functions to detect new vulnerabilities. 
The only way users might do it is to try different combinations of skipped functions, which tends to be time-consuming and ineffective. 
Worse still, skipping only coarse-grain-level code snippets on functions will not help to detect the memory leak in this case, as all the functions listed in Listing 1 are responsible for reproducing the issue. 
Fine-grained levels (e.g., statement or code block) should be supported to skip the execution of certain input-dependent loops. In summary, the above issues of \chopper make it difficult to meet the first two requirements. 

Baseline approach KLEE \cite{klee} is simply restricted by handling input-dependent loops in functions {\tt expandargv} and {\tt getopt\_long} (not shown due to page limit) before reaching the vulnerable function {\tt display\_info}, so compiling with requirements \#1 and \#2 within KLEE is hard. It may be easy to bypass these loops with certain bounds, but selecting the optimal bound to expose the issue can be challenging, as only reaching a specific number of iterations will expose the vulnerable point.

CBC \cite{yi2024compatible} bypasses the input-dependent loops successfully by pruning redundant paths that have no new contributions to branch coverage, so that requirements \#1 and \#2 are easily satisfied. 
However, as we emphasized before, achieving the best coverage has little to no correlation with the conclusion that there are no bugs.
More importantly, some subtle memory issues can only be triggered when the loop is executed with a specific number of iterations \cite{haller2013dowsing}. 
In this case, to satisfy the requirement \#3, the engine should execute a loop at least 25 times. 
To be specific, the loop inside the function {\tt getopt\_longong} scans the long string ``{\it CDE:FGHI:LM:P:RSTU:VW::Zab:defghij:lm:prstvwxz}'' one by one and returns the corresponding character iteratively. 
Only the character `i' (at position 25 among all characters) is iterated and returned, which can make the memory leak happen.
Since the loop body was covered in terms of code coverage, CBC gives a very low priority to execute the covered loop again, thus missing the detection of the memory leak issue.

\subsection{How Does \ourSol Detect it?}
\ourSol discovered the vulnerable path\footnote{\ourSol did not directly detect the memory leak, but it is the unique vulnerable path covered by \ourSol that triggers the memory leak issue.} beneficial from the new indicator for approximating the vulnerable paths and the novel search using a variant of the MCTS algorithm.
To meet the first two requirements, \ourSol skips the {\it unimportant} input-dependent loops that have no contributions to the accumulation of unsafe pointer coverage by using simulation optimization strategies (details described in Section \ref{mcts::approach:simulation}).
To explore the loops that iteratively scan the long string ``{\it CDE:FGHI:LM:P:RSTU:VW::Zab:defghij:lm:prstvwxz}'', the simulation in MCTS evaluates the reward of each iteration of the vulnerability-relevant loop, and continues the iteration when the simulation process covers new unsafe pointers (e.g., the ones in Lines 14 and 23). When the iteration reaches 25 times, the vulnerable function {\tt display\_info} is executed, and the issue is eventually detected in the function {\tt do\_display\_target} at Line 60.

\begin{table*}[t]
	\centering
	\caption{Results of CVE vulnerability reproduction (``--reverse-limit'' and ``--states-limit'' are two customized options that CBC \cite{yi2024compatible} provides for controlling the limit of traversing the dependence graph).}
	\begin{tabular}{c} 
	    \includegraphics[width=1\textwidth]{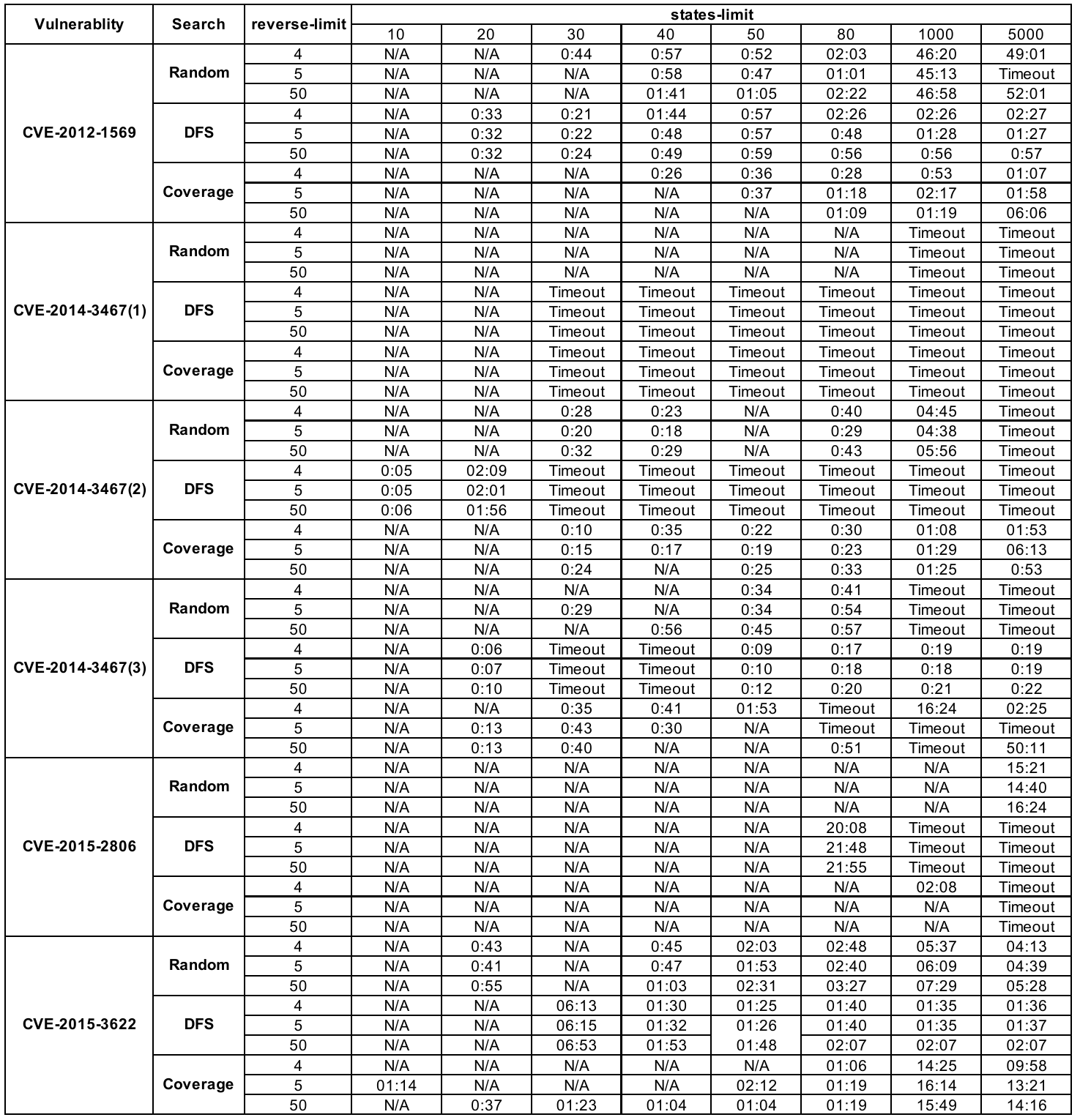}
    \end{tabular}
    \begin{tablenotes}
        \footnotesize
       \item* The time format is {\it minute:second} and ``N/A'' refers to the normal termination without reproducing the vulnerability. The index number $n$ in CVE-2014-3467($n$) represents a distinct location of the vulnerability manifested.
      \end{tablenotes}
    \label{table:cbc}
\end{table*}

\end{document}